\documentclass[journal=nalefd,manuscript=letter]{achemso}

\setkeys{acs}{doi = true}

\usepackage[version=4]{mhchem}
\usepackage{siunitx}
\usepackage[T1]{fontenc}
\usepackage{hyperref}

\usepackage{doi}

\title{Large bias-tunable magnetoresistance from spin-dependent interlayer hybridization in van der Waals antiferromagnet CrSBr-based heterostructures}

\author{Sadeed Hameed}
\altaffiliation{These authors contributed equally}
\author{Aditya Kumar}
\altaffiliation{These authors contributed equally}
\affiliation[Johannes Gutenberg University Mainz]
  {Institute of Physics, Johannes Gutenberg University Mainz, 55099 Mainz, Germany}
\author{Chengjie Yu}
\affiliation[Jiangsu University]
  {School of Physics and Electronic Engineering, Jiangsu University, Zhenjiang 212013, China}
\author{Aravind P. Balan}
\email{ap55@rice.edu}
\affiliation[Johannes Gutenberg University Mainz]
  {Institute of Physics, Johannes Gutenberg University Mainz, 55099 Mainz, Germany}
\alsoaffiliation[Rice University]
  {Department of Materials Science and NanoEngineering, Rice University, 6100 Main Street Houston, Texas 77005-1897, USA}
  \author{Xinran Wang}
\affiliation[Johannes Gutenberg University Mainz]
  {Institute of Physics, Johannes Gutenberg University Mainz, 55099 Mainz, Germany}
\author{Lichuan Zhang}
\affiliation[Jiangsu University]
  {School of Physics and Electronic Engineering, Jiangsu University, Zhenjiang 212013, China}
\author{Yuriy Mokrousov}
\affiliation[Johannes Gutenberg University Mainz]
  {Institute of Physics, Johannes Gutenberg University Mainz, 55099 Mainz, Germany}
\alsoaffiliation[Forschungszentrum J\"ulich]
  {Peter Gr\"unberg Institut and Institute for Advanced Simulation, Forschungszentrum J\"ulich and JARA, 52425 J\"ulich, Germany}
\author{Mathias Kl\"aui}
\email{klaeui@uni-mainz.de}
\affiliation[Johannes Gutenberg University Mainz]
  {Institute of Physics, Johannes Gutenberg University Mainz, 55099 Mainz, Germany}
\alsoaffiliation[Norwegian University of Science and Technology]
  {Centre for Quantum Spintronics, Department of Physics, Norwegian University of Science and Technology, Trondheim 7491, Norway}


\begin{document}

\begin{abstract}
We explore the large magnetoresistance (MR) in \ce{hBN}/few-layer-graphene/\ce{CrSBr}/few-layer-graphene heterostructures and reveal the mechanism behind its non-monotonic bias dependence. Using bias voltage and temperature as independent tuning knobs, we achieve MR up to \SI{350}{\percent} at \SI{20}{K}, characterized by symmetric M-shaped maxima around $\pm 0.5\,\mathrm{V}$. Continuous tuning of the magnetization angle $\theta$ via a hard-axis magnetic field shows that the barrier band-edge offset varies linearly with $\cos(\theta/2)$, a first-order signature of spin-dependent interlayer hybridization. This linear relationship rules out the Julli\`ere model and a spin-filter projection. We conclude that the magnetic-configuration-dependent band edge, rather than electrode spin polarization, dictates the large magnetoresistance in \ce{CrSBr} junctions.
\end{abstract}

The magnetoresistance of magnetic tunnel junctions, in which the resistance depends on the relative alignment of two ferromagnetic electrodes separated by a thin insulator, is the basis of modern non-volatile memory and magnetic-field sensing \cite{julliere1975tunneling, hirohata2022interfacial}. A distinct route to a magnetoresistive junction places the spin dependence in the barrier itself: a magnetic insulator or semiconductor between two non-magnetic electrodes presents a spin-dependent band-edge offset, so that its exchange-split conduction-band edge sets a different barrier for each spin, as shown, for instance, for \ce{EuO}/\ce{EuS} \cite{esaki1967magnetointernal, moodera1988electron}. This spin-filter junction is a different device geometry from the typical magnetic tunnel junction: the electrodes are non-magnetic, and the magnetic state of the barrier itself sets the transport. Van der Waals (vdW) heterostructures provide atomically clean interfaces that reduce spin scattering and allow for high spin polarization for this magnetic barrier route \cite{min2022tunable, gibertini2019magnetic}. A-type antiferromagnetic semiconductors are promising candidates, in which adjacent vdW layers exhibit opposite magnetizations with weaker interlayer coupling, making the barrier itself magnetically reconfigurable by an applied field.

A large magnetoresistance, strongly non-monotonic in bias voltage, has been observed across this family of van der Waals insulator junctions \cite{song2018giant, kim2018one, lan2023giant}. A non-monotonic magnetoresistance, with distinct peaks, in some cases M-shaped or sign-reversing, has been reported in \ce{CrI_3} \cite{song2018giant, song2019voltage}, \ce{NiBr_2} \cite{guo2021giant}, \ce{Fe_3GaTe_2}/\ce{GaSe} \cite{zhu2025unconventional} and \ce{Fe_3GeTe_2}/\ce{CrBr_3} \cite{wang2025large} junctions, and is also observed in the \ce{EuS} double-spin-filter junctions \cite{miao2009magnetoresistance}. The origin of this large magnetoresistance, non-monotonic with bias, however, is unclear. The same bias dependence has been interpreted as exchange-split conduction-band edges entering the bias window \cite{guo2021giant, paudel2019spin}, as resonance shifts of a layer-by-layer spin valve \cite{song2019voltage, kim2018one}, and as magnon-assisted inelastic channels \cite{klein2018probing, ghazaryan2018magnon}, and several of these interpretations remain qualitative at best \cite{miao2009magnetoresistance, wang2018very}.

Resolving the origin of the large magnetoresistance and of its bias dependence is both a practical and a conceptual necessity. The bias voltage is the operating knob of the non-volatile, multi-level vdW spintronic demonstrator devices \cite{boixconstant2024multistep, boixconstant2025programmable, chen2024twist}, and engineering the bias dependence requires understanding the underlying mechanism. For \ce{CrSBr} alone the vertical magnetoresistance has been attributed to a spin filter effect \cite{lan2023giant, cenker2023strain, cham2024spin}, or to a Mott two-current process \cite{liu2025spin}, and also to coherent band-structure tunneling distinguished from a spin valve \cite{chen2024twist}. However, the mechanism by which the magnetic configuration sets the band edge that limits the vertical transport has not been established. Consequently, the bias dependence of the magnetoresistance, and its tunability, are understood only qualitatively \cite{gibertini2019magnetic}.

\begin{figure}[!ht]
    \centering
    \includegraphics[width = \linewidth]{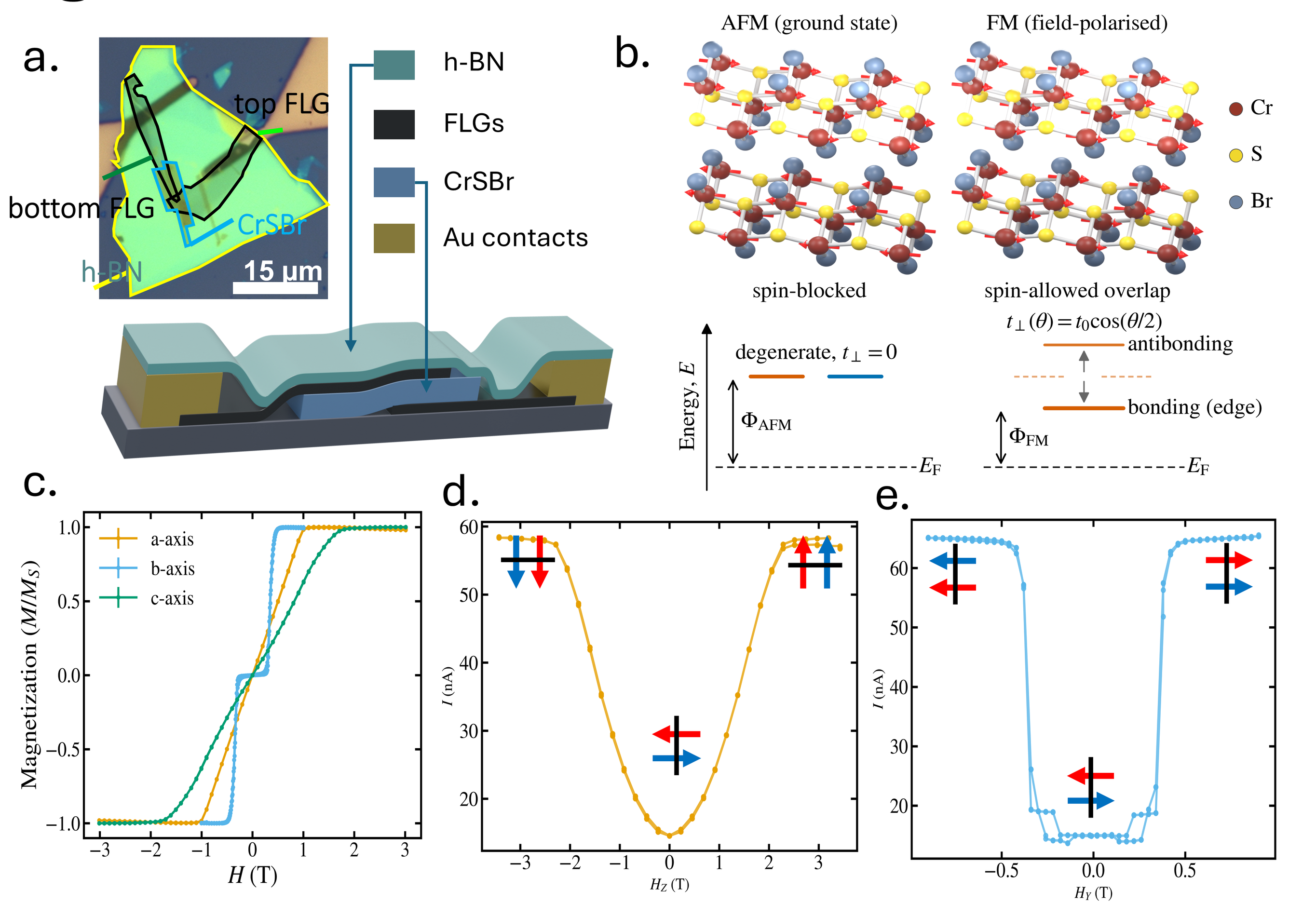}
    \caption{
    \textbf{(a)} Schematic of the \ce{hBN}/\ce{FLG}/\ce{CrSBr}/\ce{FLG} van der Waals heterostructure for vertical magnetotransport. Inset: optical micrograph of the device, with the \ce{FLG} placed on pre-patterned gold electrodes.
    \textbf{(b)} Band diagram of the junction: the band-edge offset $\Phi$ between the graphite Fermi level and the \ce{CrSBr} conduction-band edge, in the field-polarized ferromagnetic state (left, $\Phi_{\mathrm{FM}}$) and in the A-type antiferromagnetic ground state (right, $\Phi_{\mathrm{AFM}} > \Phi_{\mathrm{FM}}$).
    \textbf{(c)} SQUID magnetometry of bulk \ce{CrSBr} at base temperature: spin-flip transition near \SI{0.3}{T} along the $b$-axis (blue), and canting along the $a$-axis (yellow, saturation \SI{1}{T}) and $c$-axis (green, saturation \SI{2}{T}).
    \textbf{(d)} Magnetoresistance for sweeps of $H_{\mathrm{z}}$ along the $c$-axis, saturating near \SI{2}{T}.
    \textbf{(e)} Magnetoresistance for sweeps of $H_{\mathrm{y}}$ along the $b$-axis, with a spin-flip transition at \SI{0.3}{T}.
    }
    \label{fig:device_overview}
\end{figure}

In junctions in which an A-type antiferromagnetic semiconductor forms the barrier, the band edge depends on the interlayer magnetic configuration, presenting a higher band-edge offset $\Phi_{\mathrm{AFM}}$ in the zero-field antiferromagnetic (AFM) ground state than $\Phi_{\mathrm{FM}}$ in the field-polarized ferromagnetic (FM) state (\autoref{fig:device_overview}b). Throughout, $\Phi$ denotes the effective band-edge offset governing the transport channel being sampled. A junction whose barrier is set by such an offset converts the magnetic configuration into a large, bias-dependent magnetoresistance, making the configuration-dependent offset a candidate for the origin of the large magnetoresistance in this family. Testing this family of materials has been difficult because several effects are entangled. The prediction that the offset evolves geometrically with the angle between the magnetizations of two adjacent layers ($\theta$) has been invoked qualitatively, but tested only at the collinear AFM and FM configurations \cite{liu2025spin} rather than across the continuous angle variation in the canted states. The transport-inferred offset difference has also not been compared against an independent determination of the same quantity \cite{lin2024influence}. Moreover, the air sensitivity of these materials can often introduce interface defects and junction asymmetry that complicate transport spectroscopy and the interpretation of the results \cite{shcherbakov2018raman}.

A well-suited van der Waals magnet for such studies is chromium sulfide bromide (\ce{CrSBr}), an air-stable A-type vdW antiferromagnetic semiconductor with a N\'eel temperature $T_{\mathrm{N}} = \SI{132}{K}$ and a direct gap of $\sim\SI{1.5}{eV}$ \cite{telford2020layered, ziebel2024crsbr}; its easy ($b$) axis hosts a layer-by-layer spin-flip transition at $\sim\SI{0.3}{T}$ that switches the magnetic state of the barrier between the AFM and FM configurations \cite{telford2020layered, boixconstant2024multistep, long2023intrinsic}, while the hard ($c$) axis supports a continuous second-order canting transition that allows access to the full non-collinear interlayer-magnetization-angle continuum \cite{long2023intrinsic}. In \ce{CrSBr}, moreover, one of the candidate mechanisms has been observed independently of transport: optical spectroscopy and first-principles calculations show a spin-dependent interlayer hybridization that shifts the electronic states with the magnetic configuration, with hybridization energies on the $\sim\SI{0.1}{eV}$ scale \cite{wilson2021interlayer, heienbuttel2024quadratic}, and the vertical charge transport is known to change with the magnetic order \cite{telford2022coupling, lin2024influence}. Such a hybridization would shift the band edge, and with it the offset, continuously with the interlayer magnetization angle, a dependence that the collinear parallel and antiparallel states alone cannot distinguish from a spin-filter projection or a layer-by-layer spin valve. Measuring the transport along two axes, the continuous canting angle of magnetization and the temperature, on one device would allow one to separate these candidate mechanisms.

Here we measure the temperature and bias voltage dependence of the magnetoresistance on a single \ce{hBN}/few-layer-graphene (\ce{FLG})/\ce{CrSBr}/\ce{FLG} vertical junction, with a \SI{12}{nm} \ce{CrSBr} flake between graphite contacts (\autoref{fig:device_overview}a; thickness by atomic force microscopy, Appendix~\ref{sec:S_afm}), recording full $I$--$V$ characteristics at every field step between \SI{20}{K} and \SI{160}{K} with fields along two orthogonal magnetic axes: the easy-axis spin-flip transition fixes the AFM and FM states, and the hard-axis canting transition varies the relative angle $\theta$ between the magnetizations of adjacent layers continuously through the non-collinear continuum. From these data we perform transport spectroscopy: the normalized differential conductance $(\mathrm{d}I/\mathrm{d}V)/(I/V)$ \cite{feenstra1994tunneling, wolf2011principles} and the Fowler--Nordheim (FN) representation \cite{fowler1928electron, beebe2006transition} of the same $I$--$V$ curves identify the transport regimes along the bias and temperature axes and determine how the band-edge offset depends on the magnetic configuration.

The \ce{hBN}/\ce{FLG}/\ce{CrSBr}/\ce{FLG} heterostructure is fabricated entirely in an inert Ar atmosphere by polycarbonate-assisted dry transfer \cite{zomer2014fast, purdie2018cleaning}, with each \ce{FLG} electrode overlapping the \ce{CrSBr} at one end and a pre-patterned gold contact at the other, so that the current path runs vertically through the \ce{CrSBr} (\autoref{fig:device_overview}a). Magnetic-field sweeps with the junction current recorded at \SI{20}{K} and \SI{0.5}{V} (\autoref{fig:device_overview}d,e) reproduce the triaxial anisotropy seen in bulk \ce{CrSBr} magnetometry (\autoref{fig:device_overview}c) \cite{telford2020layered}: along the easy ($b$) axis the current jumps abruptly at $\mu_0 H = \SI{0.3}{T}$ at the layer-by-layer spin-flip transition \cite{telford2020layered, boixconstant2024multistep, long2023intrinsic}, while along the hard ($c$) axis it rises continuously and saturates near \SI{2}{T} as the magnetization cants. That agreement indicates that the junction probes the intrinsic A-type AFM order. All magnetotransport we discuss here uses fields along the easy $b$ and hard $c$ axes, and the saturation field decreases with increasing temperature as the magnetic order weakens (Appendix~\ref{sec:S_hsat}).

\begin{figure}[!ht]
    \centering
    \includegraphics[width = \linewidth]{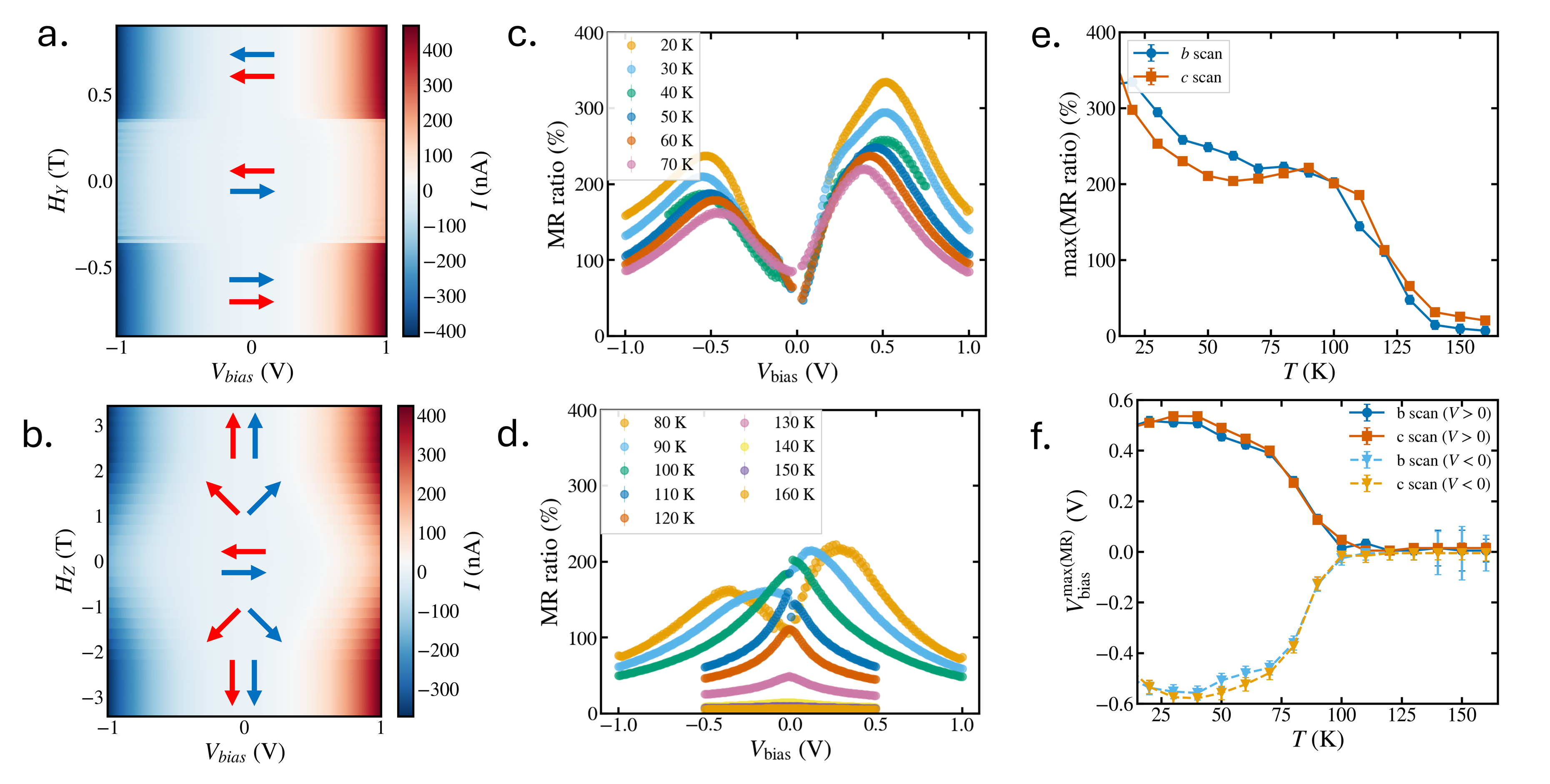}
    \caption{
    \textbf{(a)} $I$--$V$ sweeps for $H_{\mathrm{y}}$ from \SI{-0.8}{T} to \SI{0.8}{T} along the $b$-axis.
    \textbf{(b)} $I$--$V$ sweeps for $H_{\mathrm{z}}$ from \SI{-3.5}{T} to \SI{3.5}{T} along the $c$-axis.
    \textbf{(c,d)} Bias dependence of the MR ratio between the ferromagnetic and antiferromagnetic states for field along the $b$-axis, from \SI{20}{K} to \SI{70}{K} (c) and from \SI{80}{K} to \SI{160}{K} (d). The $c$-axis counterpart is given in Appendix~\ref{sec:S_baxis}.
    \textbf{(e)} Maximum magnetoresistance versus temperature, $b$- and $c$-axis.
    \textbf{(f)} Positive and negative bias of maximum magnetoresistance versus temperature, $b$- and $c$-axis.
    In (c)--(f) the ferromagnetic state is the fully field-polarized state, with $I_{\mathrm{FM}}$ averaged over $|\mu_{0}H_{\mathrm{y}}| > \SI{0.5}{T}$ ($b$-axis) and $|\mu_{0}H_{\mathrm{z}}| > \SI{2.5}{T}$ ($c$-axis), and $I_{\mathrm{AFM}}$ taken near zero field.
    }
    \label{fig:TMR_bias_temperature}
\end{figure}

$I$--$V$ characteristics were recorded between \SI{-1}{V} and \SI{+1}{V} during the field sweeps along the $b$-axis (\autoref{fig:TMR_bias_temperature}a) and the $c$-axis (\autoref{fig:TMR_bias_temperature}b), and the magnetoresistance is defined as
\begin{align}
    \mathrm{MR}(V, T) = \frac{I_{\mathrm{FM}}(V, T) - I_{\mathrm{AFM}}(V, T)}{I_{\mathrm{AFM}}(V, T)} ,
    \label{eq: MR}
\end{align}
where $I_{\mathrm{FM}}$ ($I_{\mathrm{AFM}}$) is the current in the field-polarized FM (zero-field AFM) state. The FM state is the fully saturated state, reached above the $b$-axis spin-flip transition and above the $c$-axis saturation; $I_{\mathrm{FM}}$ is averaged over $|\mu_{0}H_{\mathrm{y}}| > \SI{0.5}{T}$ along the $b$-axis and over $|\mu_{0}H_{\mathrm{z}}| > \SI{2.5}{T}$ along the $c$-axis, and $I_{\mathrm{AFM}}$ near zero field. At \SI{20}{K}, $\mathrm{MR}(V)$ is non-monotonic and M-shaped for field along the $b$-axis (\autoref{fig:TMR_bias_temperature}c), with two peaks near $V \approx \pm\SI{0.5}{V}$ reaching $\sim\SI{350}{\percent}$. The modest asymmetry between positive and negative bias could arise from an asymmetric Schottky barrier at the structurally inequivalent top and bottom \ce{FLG}/\ce{CrSBr} interfaces (Appendix~\ref{sec:S_pinning}); a comparable bias asymmetry is observed in \ce{CrI_3} \cite{song2018giant, song2019voltage} and \ce{EuS} \cite{miao2009magnetoresistance} junctions, and the M-shape itself persists for a symmetric junction. The $b$-axis sweeps yield a similar MR ratio and the same temperature and bias voltage dependence as the $c$-axis (\autoref{fig:TMR_bias_temperature}c--f; Appendix~\ref{sec:S_baxis}). With increasing temperature the two peaks move to smaller bias and merge near \SI{90}{K}, above which $\mathrm{MR}(V)$ becomes monotonic in the magnitude of the bias voltage, and the maximum magnetoresistance falls below \SI{100}{\percent} near \SI{100}{K} before the signal vanishes at $T_{\mathrm{N}}$ (\autoref{fig:TMR_bias_temperature}c--f). To determine the origin of the M-shape and of its disappearance we turn to transport spectroscopy of the same $I$--$V$ data.

Vertical transport in \ce{CrSBr} is hopping-dominated and thermally activated across the 5.4 to \SI{83}{nm} thickness range studied by Lin \emph{et al.}~\cite{lin2024influence}, and for our \SI{12}{nm} thickness, coherent elastic tunneling through the entire thickness is excluded and the current is limited by the band-edge barrier of the \ce{CrSBr} rather than by a single interface. The bias-scale features of the $I$--$V$ curves therefore reflect the band edge of the barrier material itself, and we analyze them with two complementary representations of the same data. The FN representation $\ln(|I|/V^{2})$ versus $1/V$ shows a minimum at the transition voltage $V_{T}$, the bias at which transport crosses from low-bias hopping to field emission over the barrier, and $eV_{T}$ locates the band-edge offset independent of the barrier width and effective mass \cite{simmons1963generalized, beebe2006transition, trouwborst2011transition, baldea2012transition}, whereas the normalized differential conductance $(\mathrm{d}I/\mathrm{d}V)/(I/V)$ develops a maximum at a bias $V_{\mathrm{peak}}$ that can be followed at every field step where the FN minimum is not resolved. Both representations are used only as phenomenological descriptions. The $c$-axis bandwidth is just a few meV \cite{lin2024influence}, so the free-electron dispersion assumed by textbook FN theory does not apply. Likewise, $V_{\mathrm{peak}}$ does not represent the band-edge offset alone: it also depends on how the bias voltage is divided across the device, the electrodes' density of states, and the shape of the tunneling transmission. What each yields robustly is the position of the features along the bias axis and their differences with field, where the conversion to an energy carrying a common factor of order unity, $\Phi = eV_{\mathrm{peak}}/c$, whose calibration we defer to the comparison with calculated energies below.

\begin{figure}[!ht]
    \centering
    \includegraphics[width = \linewidth]{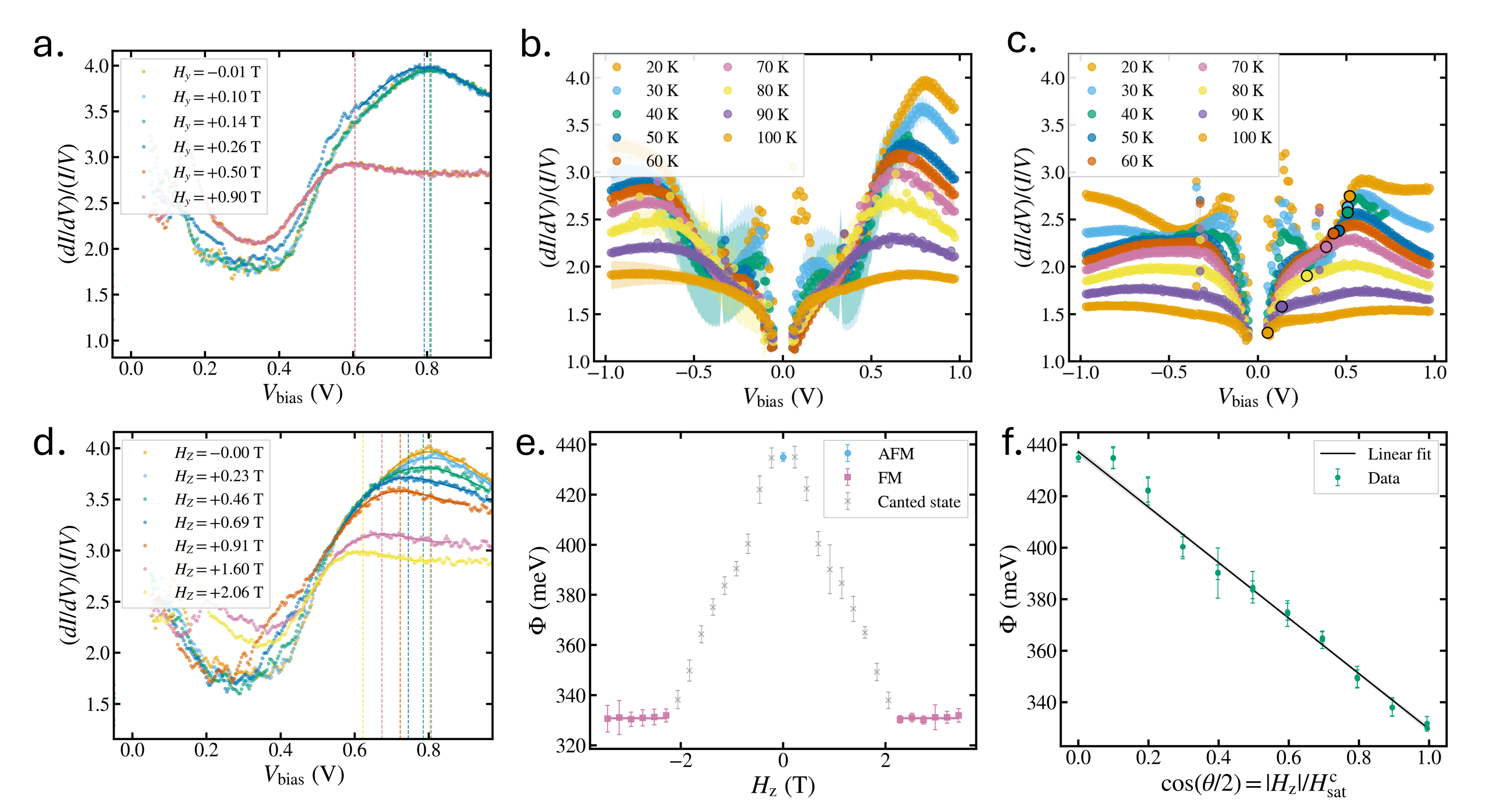}
    \caption{
    \textbf{(a)} Normalized differential conductance $(\mathrm{d}I/\mathrm{d}V)/(I/V)$ at \SI{20}{K} along the easy $b$-axis, for four antiferromagnetically aligned fields ($H_{\mathrm{y}} = -0.01, +0.10, +0.14, +0.26$~T, of which the first three lie inside the pristine AFM window $|H_{\mathrm{y}}| < \SI{0.2}{T}$ used for the weighted fit) and two in the saturated FM window ($H_{\mathrm{y}} = +0.50, +0.90$~T). Points are data, solid lines Gaussian-plus-linear-background fits; vertical dashed lines mark the fitted peak positions $V_{\mathrm{peak}}^{\mathrm{FM}} \approx \SI{0.62}{V}$ and $V_{\mathrm{peak}}^{\mathrm{AFM}} \approx \SI{0.80}{V}$.
    \textbf{(b,c)} Normalized differential conductance from \SI{20}{K} to \SI{100}{K} in the AFM (b) and FM (c) states, for fields along both the easy $b$- and hard $c$-axes.
    \textbf{(d)} Normalized differential conductance at \SI{20}{K} for fields $H_{\mathrm{z}}$ along the hard $c$-axis, from $H_{\mathrm{z}} = 0$ (AFM) to $H_{\mathrm{z}} \approx +2.06$~T ($h \approx 0.94$, near saturation).
    \textbf{(e)} Band-edge offset $\Phi(H_{\mathrm{z}}) = eV_{\mathrm{peak}}/c$ from the per-curve fits, $c$-axis, with $c = 1.85$ (measured in the AFM state and assumed configuration-independent, Appendix~\ref{sec:S_fn}); the AFM and FM windows and the intermediate-angle trajectory are marked.
    \textbf{(f)} $\Phi$ versus $\cos(\theta/2) = |H_{\mathrm{z}}|/H_{\mathrm{sat}}$, $c$-axis (points), with an inverse-variance-weighted linear fit (solid line, gray 1-$\sigma$ band). The comparison with the $\sin^{2}(\theta/2)$ form is given in Appendix~\ref{sec:S_model}.
    }
    \label{fig:spectroscopy_and_canting}
\end{figure}

At \SI{20}{K} the normalized conductance peaks at $V_{\mathrm{peak}} \approx \SI{0.62}{V}$ in the FM state and $V_{\mathrm{peak}} \approx \SI{0.80}{V}$ in the AFM state (\autoref{fig:spectroscopy_and_canting}a). $V_{\mathrm{peak}}$ is extracted from each $I$--$V$ curve by fitting a Gaussian plus a linear background to the normalized differential conductance, and inverse-variance-weighted fits over the saturated FM ($|H_{\mathrm{y}}| > \SI{0.5}{T}$) and pristine AFM ($|H_{\mathrm{y}}| < \SI{0.2}{T}$) windows yield peak positions differing by $\Delta V_{\mathrm{peak}} = (0.193 \pm 0.003)\,\mathrm{V}$. In the FN representation the AFM state passes through a transition-voltage minimum at $V_{T}^{\mathrm{AFM}} \approx \SI{0.44}{V}$ (Appendix~\ref{sec:S_fn}). These voltages bound the channel being probed: $eV_{T}^{\mathrm{AFM}} \approx \SI{0.44}{eV}$ is far below the $\sim\SI{1.5}{eV}$ optical gap \cite{telford2020layered, ziebel2024crsbr}, and the complementary hole barrier ($\approx \SI{1.1}{eV}$) lies outside the $\pm\SI{1}{V}$ bias window, so electron injection toward the conduction-band edge is the only channel probed. This configuration-dependent barrier is the origin of the M-shaped bias dependence of $\mathrm{MR}(V)$ (\autoref{fig:TMR_bias_temperature}c): at low bias both currents are small and the FM/AFM ratio is modest, but as the bias approaches $V_{\mathrm{peak}}^{\mathrm{FM}}$ the field-polarized state crosses into field emission over its lower barrier while the AFM state, with the higher barrier, is still suppressed, so the signal rises to its maximum at $V^{\ast} \approx \SI{0.5}{V}$, just below $V_{\mathrm{peak}}^{\mathrm{FM}}$; at still higher bias the AFM state in turn enters field emission, both channels conduct, and the ratio falls, producing the outer arm of the M-shape.

The AFM and FM states alone do not fix the mechanism, and the hard-$c$-axis canting transition provides a geometric test across the non-collinear continuum. Balancing interlayer exchange and Zeeman energy gives a relative interlayer angle $\theta(H_{\mathrm{z}}) = 2\arccos(|H_{\mathrm{z}}|/H_{\mathrm{sat}})$ between the layer magnetizations, so $\cos(\theta/2) = |H_{\mathrm{z}}|/H_{\mathrm{sat}} \equiv h$ runs from $0$ (AFM) to $1$ (FM-aligned). The band-edge-offset peak yields $\Phi(H_{\mathrm{z}}) = eV_{\mathrm{peak}}/c$, which shifts continuously between the AFM and FM states and is symmetric about zero field (\autoref{fig:spectroscopy_and_canting}d,e), with $c$-axis AFM and FM values matching the $b$-axis values to within \SI{2}{} to \SI{5}{meV}. A static interface feature would reproduce neither that continuous shift nor its symmetry. Plotting $\Phi$ against $h$ (\autoref{fig:spectroscopy_and_canting}f), the data follow a linear law, which is favored over the quadratic $\sin^{2}(\theta/2) = 1 - h^{2}$ form by the per-temperature reduced $\chi^2$; the two laws are compared in detail in Appendix~\ref{sec:S_model}.

The linear dependence is a signature of an offset set by a band-edge energy. In the AFM configuration the conduction-band-edge states of adjacent layers are degenerate, and the spin-dependent interlayer coupling $t_{\perp}(\theta) = t_{0}\cos(\theta/2)$ \cite{wilson2021interlayer, heienbuttel2024quadratic} splits them in first order, lowering the band edge, and with it the offset, in proportion to $|t_{\perp}|$, giving
\begin{align}
    \Phi(\theta) = \Phi_{\mathrm{AFM}} - \left(\Phi_{\mathrm{AFM}} - \Phi_{\mathrm{FM}}\right)\cos(\theta/2) ,
    \label{eq: hybridization}
\end{align}
linear in $h = \cos(\theta/2)$. We therefore attribute the configuration dependence of the offset to spin-dependent interlayer hybridization rather than to a spin-filter projection or an interface dipole.

That identification is independent of the bias-to-energy conversion: the linear dependence on $\cos(\theta/2)$, its symmetry about zero field, and the agreement between the two axes are all unaffected by a common multiplicative factor. The energy scale corroborates it. Both observables are available in the AFM alignment, which fixes the conversion: the transition voltage yields $\Phi_{\mathrm{AFM}} = eV_{T}^{\mathrm{AFM}} \approx \SI{0.44}{eV}$, calibrating $c \equiv V_{\mathrm{peak}}^{\mathrm{AFM}}/V_{T}^{\mathrm{AFM}} = 1.85$, which is empirical and assumed configuration-independent rather than justified from transition-voltage-spectroscopy theory. The field-polarized state then presents the lower offset, $\Phi_{\mathrm{FM}} = eV_{\mathrm{peak}}^{\mathrm{FM}}/c \approx \SI{0.33}{eV}$, a difference $\Phi_{\mathrm{AFM}} - \Phi_{\mathrm{FM}} \approx \SI{0.11}{eV}$. Additive, state-independent contributions (image-charge lowering, contact work-function asymmetry) cancel in that difference, while the calibration factor rescales it, so we treat the absolute values as indicative and compare on scale only: the measured $\sim\SI{0.11}{eV}$ lies on the same $\sim\SI{0.1}{eV}$ scale as the calculated spin-dependent interlayer hybridization energies \cite{wilson2021interlayer, heienbuttel2024quadratic}.

These results separate the band-edge-offset picture from the two pictures that dominate the 2D-magnet tunneling literature. In a layer-by-layer spin valve, the magnetoresistance is a conductance ratio set by reversing the A-type stack between antiparallel and parallel alignment \cite{song2018giant, song2019voltage, gibertini2019magnetic}; here the measured quantity is a band-edge-offset energy that shifts continuously with the relative angle $\theta$ as $\cos(\theta/2)$ (\autoref{fig:spectroscopy_and_canting}f) and enters the field-emission regime depending on magnetic state and bias voltage, the signature of a configuration-dependent band-edge barrier, as reported in \ce{EuS} \cite{esaki1967magnetointernal} and \ce{CrBr_3} \cite{wang2021magnetization}. Magnon-assisted inelastic tunneling \cite{ghazaryan2018magnon, klein2018probing} cannot produce a feature at the \SI{0.6}{} to \SI{0.8}{V} field-emission onset, two orders of magnitude above the meV magnon scale. Tunneling anisotropic magnetoresistance would appear as a difference between the field-saturated responses along the two crystal axes; the $b$- and $c$-axis sweeps yield MR ratios that agree to within \SI{20}{\percent} and the same bias dependence (\autoref{fig:TMR_bias_temperature}c--f, Appendix~\ref{sec:S_baxis}), so we find no evidence for a contribution comparable to the magnetoresistance itself, consistent with the small spin-orbit coupling of the near-gap states of \ce{CrSBr} \cite{heienbuttel2024quadratic}. The linear-in-$\cos(\theta/2)$ shape additionally excludes the Julli\`ere model for TMR, which would lead to a quadratic $1 - h^{2}$ dependence \cite{julliere1975tunneling} and which is not expected to apply to a junction with non-magnetic electrodes.

\begin{figure}[!ht]
    \centering
    \includegraphics[width = \linewidth]{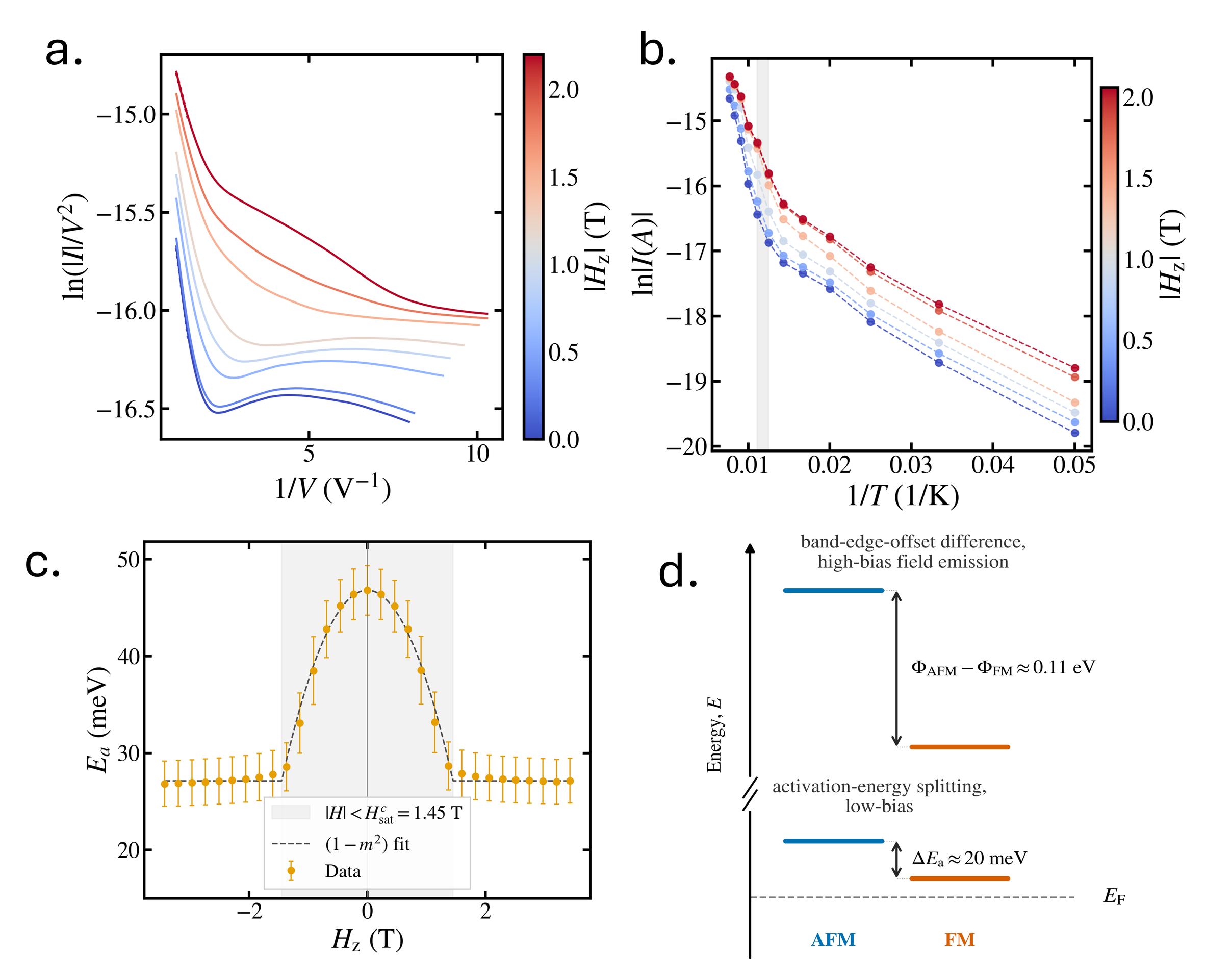}
    \caption{
    \textbf{Two transport regimes, separated by a crossover near \SI{90}{K} ($\sim 0.7\,T_{\mathrm{N}}$).}
    \textbf{(a)} Fowler--Nordheim representation $\ln(|I|/V^{2})$ versus $1/V$ at \SI{20}{K} for fields $H_{\mathrm{z}}$ along the $c$-axis, from the AFM state ($H_{\mathrm{z}}=0$) to the FM state (color scale); Appendix~\ref{sec:S_fn}.
    \textbf{(b)} Arrhenius plot $\ln|I|$ versus $1/T$ in the AFM and FM states, $c$-axis, at $V_{\mathrm{probe}} = \SI{0.2}{V}$. Below \SI{90}{K} the slopes coincide; above \SI{90}{K} they split, with $E_{\mathrm{a}}^{\mathrm{AFM}} \approx \SI{47}{meV}$, $E_{\mathrm{a}}^{\mathrm{FM}} \approx \SI{27}{meV}$ and $\Delta E_{\mathrm{a}} \approx \SI{20}{meV}$. Error bars, propagated from the measured currents, are smaller than the markers. Fit-form comparison: Appendix~\ref{sec:S_arrhenius}.
    \textbf{(c)} Activation energy $E_{\mathrm{a}}(H_{\mathrm{z}})$ above \SI{90}{K} for a probe bias of \SI{0.2}{V}, $c$-axis, with a fit to $E_{\mathrm{a}} = E_{\mathrm{a}}^{\mathrm{FM}} + (E_{\mathrm{a}}^{\mathrm{AFM}} - E_{\mathrm{a}}^{\mathrm{FM}})(1 - m^{2})$, $m = |H_{\mathrm{z}}|/H_{\mathrm{sat}}$.
    \textbf{(d)} Energy-level diagram of the two AFM-to-FM splittings: $\sim\SI{20}{meV}$ at the band edge from the low-bias activation of (b), and $\sim\SI{0.11}{eV}$ from high-bias field emission at energies of order $\Phi$ above $E_F$.
    }
    \label{fig:two_regimes}
\end{figure}

The temperature dependence of the low-bias current reveals two regimes separated by a crossover near \SI{90}{K}, $\sim 0.7\,T_{\mathrm{N}}$ (\autoref{fig:two_regimes}b), at which the differential-conductance peaks (\autoref{fig:spectroscopy_and_canting}b,c; Appendix~\ref{sec:S_dgdv}) also merge. Below \SI{90}{K} the current is variable-range hopping (VRH): among the standard forms the Efros--Shklovskii law $\sigma \propto \exp[-(T_{0}/T)^{1/2}]$ \cite{efros1975coulomb}, with $T_{0}$ set by the localization length and the density of states rather than a gap, is preferred over Mott and Arrhenius descriptions (Appendix~\ref{sec:S_arrhenius}), the signature of a Coulomb gap in disorder-localized band-tail states \cite{meir1996universal, huang2022conductivity}. In this regime the apparent activation energy is small ($\sim\SI{6.5}{meV}$) and field-independent: the AFM and FM states share the same slope, $\Delta E_{\mathrm{a}} \approx 0$. Above \SI{90}{K} the current is thermally activated over the conduction-band edge, $I \propto \exp(-E_{\mathrm{a}}/k_{B}T)$, and the activation energy becomes field-dependent, with $E_{\mathrm{a}}^{\mathrm{AFM}} \approx \SI{47}{meV}$, $E_{\mathrm{a}}^{\mathrm{FM}} \approx \SI{27}{meV}$ and an AFM-to-FM splitting $\Delta E_{\mathrm{a}} \approx \SI{20}{meV}$ (\autoref{fig:two_regimes}b).

Within a thermally activated band-edge model, this $\sim\SI{20}{meV}$ difference corresponds to the AFM-to-FM shift of the conduction-band edge, consistent with an exchange splitting there, and it agrees with the value Lin \emph{et al.}\ report for vertical hopping transport in \ce{CrSBr} \cite{lin2024influence}. Below \SI{90}{K} the carriers do not reach the conduction-band edge but hop between localized band-tail states, so the small $\sim\SI{6.5}{meV}$ slope is a hopping scale rather than a band-edge energy, and it is nearly the same for both magnetic states because the hopping path does not involve the band edge; the magnetic state modifies the accessible hopping path rather than the activation slope, as reported for the linear-response regime \cite{lin2024influence} and compared with our finite-bias data in Appendix~\ref{sec:S_arrhenius}. Above \SI{90}{K} the carriers are thermally excited from the band-tail states over the band edge, so $E_{\mathrm{a}} = E_{\mathrm{CBM}} - E_F$ measures the band-edge position and its AFM-to-FM difference is the exchange splitting at the band edge.

The FN evolution (\autoref{fig:two_regimes}a) and the continuous shift of the differential-conductance peak (\autoref{fig:spectroscopy_and_canting}d,e) are two views of the same change of the interlayer hybridization with $\theta$: the enhanced interlayer hopping of the spin-aligned state raises the low-bias current while the lowered band-edge offset moves the field-emission onset to lower bias, so the transition-voltage minimum of the AFM state evolves continuously into a gradual change of slope toward the FM alignment case.

The two field-dependence shapes follow from the same relative angle $\theta$, entering at different orders. In the low-temperature field-emission channel, the offset follows the interlayer-hybridization amplitude $\cos(\theta/2) = h$, first order in $t_{\perp}$, and is linear in $h$ (\autoref{fig:spectroscopy_and_canting}f); in the high-temperature activated channel, the thermally averaged activation energy follows the interlayer spin correlation $\langle\mathbf{S}_{1}\!\cdot\!\mathbf{S}_{2}\rangle \propto 2m^{2}-1$, with $m = M_{\mathrm{z}}/M_{\mathrm{sat}} = h$ below saturation, giving the quadratic $1 - m^{2}$ form of the spin-disorder model of Mauger and Godart \cite{mauger1986magnetic}, which the field-dependent activation energy follows (\autoref{fig:two_regimes}c) across the full probe-bias range (Appendix~\ref{sec:S_arrhenius}). The two probes therefore yield two AFM-to-FM splittings of common magnetic origin but different size (\autoref{fig:two_regimes}d): $\sim\SI{20}{meV}$, sampled at the band edge by low-bias activation, and the band-edge-offset difference of $\sim\SI{0.11}{eV}$, sampled by high-bias field emission at energies of order $\Phi$ above the Fermi level. The effective splitting thus grows, by a factor of about five, from the band edge toward the energies that field emission samples. The crossover near $0.7\,T_{\mathrm{N}}$ appears in three independent measurements: the field-emission slope changes sign, the differential-conductance peak disappears, and the activation energy turns field-dependent (\autoref{fig:two_regimes}b,c).

The crossover also explains the disappearance of the M-shaped $\mathrm{MR}(V)$. The M-shape requires the configuration-dependent offset to act as a field-emission barrier, with the FM state crossing into field emission at a lower bias than the AFM state. Above the crossover, transport is thermally activated over the band edge irrespective of bias, so the abrupt bias-dependent onset that defines the M-shape is lost, and the differential-conductance peaks fade into the background, so that the maximum magnetoresistance falls below \SI{100}{\percent} near \SI{100}{K} (\autoref{fig:TMR_bias_temperature}e) and vanishes at $T_{\mathrm{N}}$. Spectral broadening and magnon-assisted inelastic channels could further reduce the signal \cite{ghazaryan2018magnon, zhang1997quenching}, although we do not model them here.

In summary, from the temperature and bias voltage dependence of the magnetoresistance of a single \ce{CrSBr} junction, we have identified spin-dependent interlayer hybridization, which shifts the band edge with the magnetic configuration, as the origin of the large magnetoresistance, of its M-shaped bias dependence, and of its disappearance with increasing temperature. The band-edge offset varies linearly with the interlayer-hybridization amplitude $\cos(\theta/2)$ as the field rotates the layer magnetizations, with $\Phi_{\mathrm{AFM}} - \Phi_{\mathrm{FM}} \approx \SI{0.11}{eV}$ at \SI{20}{K}, on the $\sim\SI{0.1}{eV}$ scale of the calculated interlayer hybridization \cite{wilson2021interlayer, heienbuttel2024quadratic}, which separates the hybridization mechanism from a layer-by-layer spin valve, magnon-assisted tunneling, and the Julli\`ere model for TMR. The temperature dependence follows from two transport regimes separated by a crossover near $0.7\,T_{\mathrm{N}}$: variable-range hopping with a field-independent activation energy below it, and thermally activated transport over the conduction-band edge above it, with an AFM-to-FM splitting of the activation energy of $\sim\SI{20}{meV}$ consistent with prior vertical-transport work \cite{lin2024influence}. Why the effective splitting grows from $\sim\SI{20}{meV}$ at the band edge to $\sim\SI{0.11}{eV}$ at the energies sampled by field emission remains an open question.

The magnetoresistance of this van der Waals antiferromagnetic junction is therefore set by a magnetic-configuration-dependent band edge rather than by electrode spin polarization. The bias voltage acts as an energy-resolved probe of the barrier electronic structure, and the resistance change can be programmed by where the bias sits relative to the two band edges, offering a template for reading out magnetic order in other semiconducting van der Waals antiferromagnets.

\section{Methods}

\subsection{Device fabrication}
Bulk graphite, \ce{hBN}, and \ce{CrSBr} crystals (HQ Graphene) were exfoliated; \ce{CrSBr} in an argon glovebox (MBRAUN LABstar) with \ce{H_2O} and \ce{O_2} below \SI{0.5}{ppm}, and \ce{FLG} by heat-assisted exfoliation \cite{huang2015reliable}. Flakes were identified by optical microscopy and characterized by atomic force microscopy and Raman spectroscopy, with the \ce{CrSBr} thickness estimated by optical contrast. Devices were assembled in the argon glovebox by polycarbonate-assisted dry transfer \cite{zomer2014fast, purdie2018cleaning} of the \ce{hBN}/\ce{Graphene}/\ce{CrSBr}/\ce{Graphene} stack; pre-patterned \ce{Au}(\SI{10}{nm})/\ce{Cr}(\SI{2}{nm}) contacts were dry-transferred onto the few-layer-graphene electrodes, and the \ce{hBN} capping that keeps the air-sensitive \ce{CrSBr} encapsulated was completed in the inert atmosphere.

\subsection{Magnetotransport measurements}
Measurements used a 3D vector cryostat (down to \SI{1.6}{K}, up to \SI{5}{T} vertical and \SI{1}{T} horizontal). A Keithley 2400 SourceMeter applied a DC bias and measured the current in a two-terminal configuration; a Gaussian smoothing filter ($\sigma = 1.5$) reduced low-bias noise. $I$--$V$ curves were recorded at each field step; $I_{\mathrm{FM}}$ was averaged over $|H| > \SI{0.5}{T}$ ($b$-axis) and $|H| > \SI{2.5}{T}$ ($c$-axis), and $I_{\mathrm{AFM}}$ near zero field. The in-plane $b$-axis was identified by angular-dependent resistance at \SI{10}{K} and \SI{0.5}{T}.

\subsection{Transport spectroscopy}
Differential conductance was obtained by differentiating the smoothed $I(V)$ numerically on the bias grid $V_{i}$, using the central difference
\begin{align}
    \left.\frac{\mathrm{d}I}{\mathrm{d}V}\right|_{V_{i}} = \frac{I_{i+1} - I_{i-1}}{V_{i+1} - V_{i-1}} ,
    \label{eq:central_difference}
\end{align}
which is second-order accurate on the uniform grid used here, and normalizing as $(\mathrm{d}I/\mathrm{d}V)/(I/V)$ \cite{feenstra1994tunneling}. The second derivative was obtained by applying \autoref{eq:central_difference} again to $\mathrm{d}I/\mathrm{d}V$. The conductance maximum $V_{\mathrm{peak}}$ was extracted per curve within the band-edge window \SI{0.40}{} to \SI{0.98}{V} by a Gaussian-plus-linear-background fit. The AFM and FM values were obtained from inverse-variance-weighted means of the per-curve $V_{\mathrm{peak}}$ over the corresponding field windows.

\begin{acknowledgement}
We acknowledge funding from the EU Marie-Curie Postdoctoral Fellowship ExBiaVdW (Grant ID: 101068014), Deutsche Forschungsgemeinschaft (DFG, German Research Foundation) -- Spin$+$X TRR 173--268565370 (Projects No. A01, A03, A11, B02, B15 and A12), DFG Project No. 358671374, Graduate School of Excellence Materials Science in Mainz (MAINZ) GSC 266, and the Research Council of Norway (Centre for Quantum Spintronics - QuSpin No. 262633). Y.M. acknowledges funding by the DFG in the framework of TRR 288/2 - 422213477 (Project B06), TRR 173/3 - 268565370 (project A11) and Jülich Supercomputing Centre for providing computational resources under project jiff40. L.-C. Z. acknowledges financial support from the National Natural Science Foundation of China (Grant No. 12404051).
\end{acknowledgement}


\clearpage
\setcounter{section}{0}
\setcounter{figure}{0}
\setcounter{table}{0}
\setcounter{equation}{0}
\renewcommand{\thesection}{S\arabic{section}}
\renewcommand{\thesubsection}{S\arabic{section}.\arabic{subsection}}
\renewcommand{\thefigure}{S\arabic{figure}}
\renewcommand{\thetable}{S\arabic{table}}
\renewcommand{\theequation}{S\arabic{equation}}

\begin{center}
{\Large\bfseries Supporting Information}
\end{center}

This Supporting Information collects the data and the fits that support the main text: device topography (Appendix~\ref{sec:S_afm}), the complementary-field-axis measurements (Appendix~\ref{sec:S_baxis}), the angular-law model comparison (Appendix~\ref{sec:S_model}), the bias-to-energy calibration (Appendix~\ref{sec:S_pinning}), the saturation field (Appendix~\ref{sec:S_hsat}), the differential-conductance temperature series (Appendix~\ref{sec:S_dgdv}), the Fowler--Nordheim analysis (Appendix~\ref{sec:S_fn}), the activated-transport and variable-range-hopping analysis (Appendix~\ref{sec:S_arrhenius}), and the first-principles calculations (Appendix~\ref{sec:S_dft}).

\section{Device topography}
\label{sec:S_afm}

\begin{figure}[h]
    \centering
    \includegraphics[width = 0.55\linewidth]{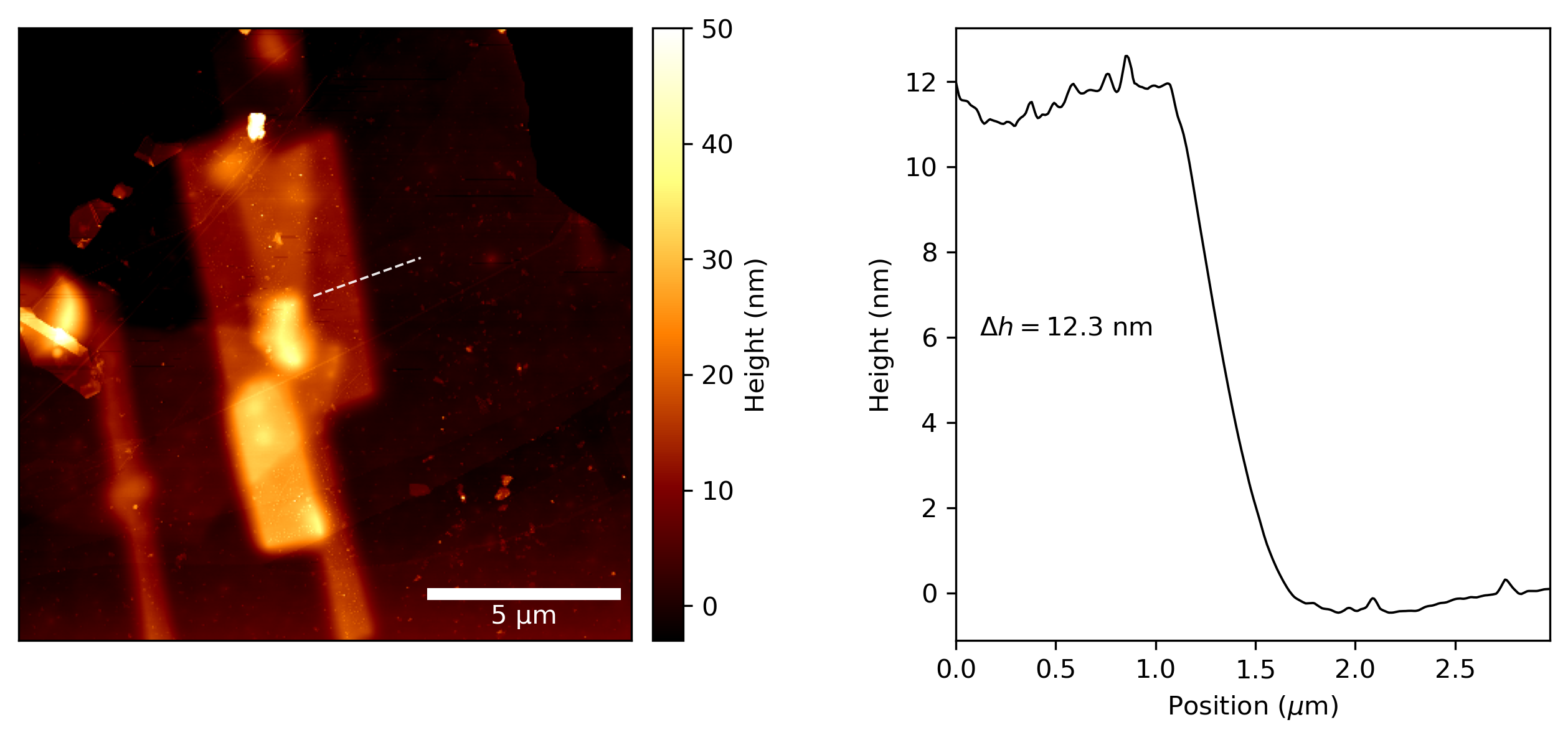}
    \caption{
    \textbf{Atomic force microscopy topography of the device region.} The \ce{CrSBr} flake and the few-layer-graphene electrodes are visible; the height is referenced to the substrate. Scale bar \SI{5}{\micro\meter}.
    }
    \label{fig:S_afm}
\end{figure}

\autoref{fig:S_afm} shows an atomic force microscopy topograph of the device region, resolving the \ce{CrSBr} flake and the few-layer-graphene electrodes of the vertical junction.

\section{Complementary-axis AFM and FM measurements}
\label{sec:S_baxis}

For each main-text panel, the measurement along the complementary field axis is collected here. The two axes give the same behavior in the two collinear states.

\begin{figure}[h]
    \centering
    \includegraphics[width = \linewidth]{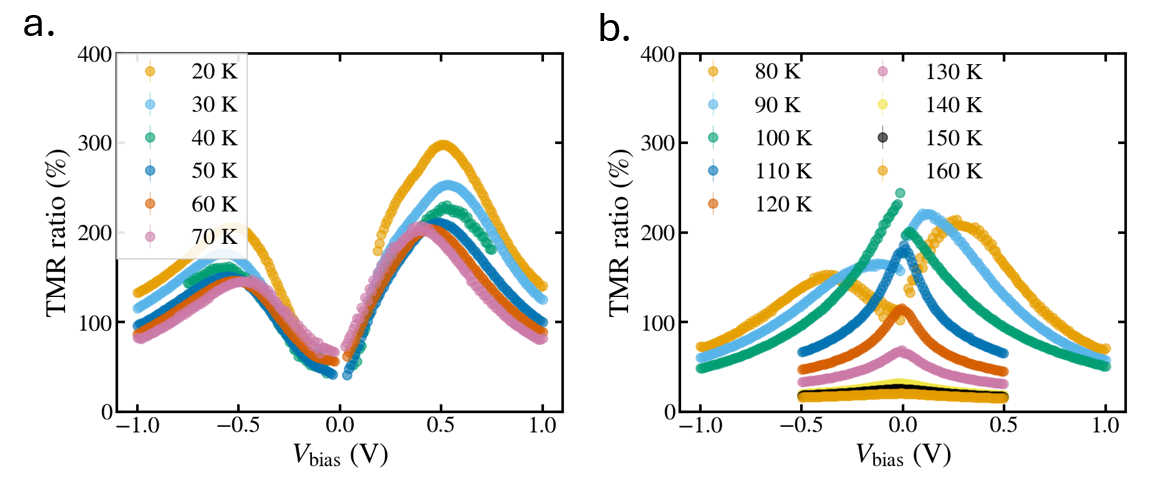}
    \caption{
    \textbf{Bias dependence of the hard-axis ($c$-axis) magnetoresistance at fixed temperature.}
    \textbf{(a)} \SI{20}{} to \SI{70}{K}. \textbf{(b)} \SI{80}{} to \SI{160}{K}. The ferromagnetic state is the fully field-polarized state, with $I_{\mathrm{FM}}$ averaged over $|\mu_{0}H_{\mathrm{z}}| > \SI{2.5}{T}$ and $I_{\mathrm{AFM}}$ taken near zero field. The M-shape of $\mathrm{MR}(V)$ is lost as $T \to T_{\mathrm{N}}$; counterpart of \autoref{fig:TMR_bias_temperature}c,d, which is measured along the easy $b$-axis.
    }
    \label{fig:S_caxis_TMR}
\end{figure}

\begin{figure}[h]
    \centering
    \includegraphics[width = \linewidth]{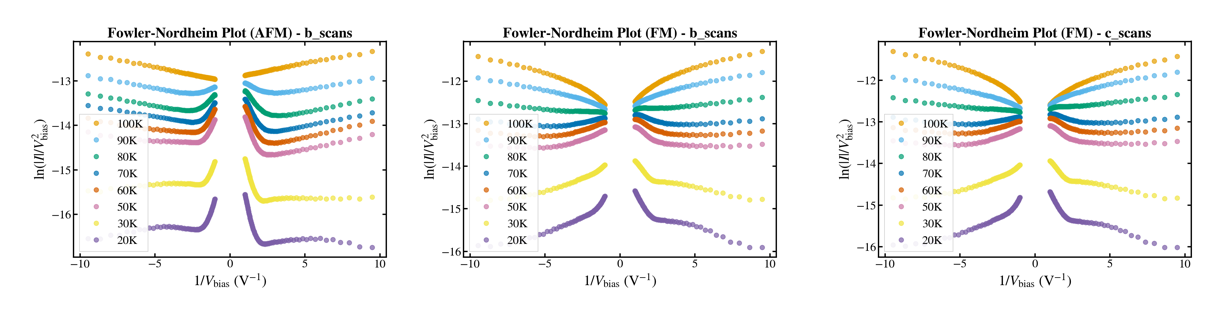}
    \caption{
    \textbf{Fowler--Nordheim representation $\ln(|I|/V^{2})$ versus $1/V$ in the AFM and FM states.}
    \textbf{(a)} AFM state ($b$-axis, zero field). \textbf{(b)} Field-polarized FM state along the easy $b$-axis. \textbf{(c)} Field-polarized FM state along the hard $c$-axis. The $b$-axis AFM and FM states show the same field-emission character as the $c$-axis; counterpart of \autoref{fig:two_regimes}a.
    }
    \label{fig:S_baxis_FN}
\end{figure}

\begin{figure}[h]
    \centering
    \includegraphics[width = 0.7\linewidth]{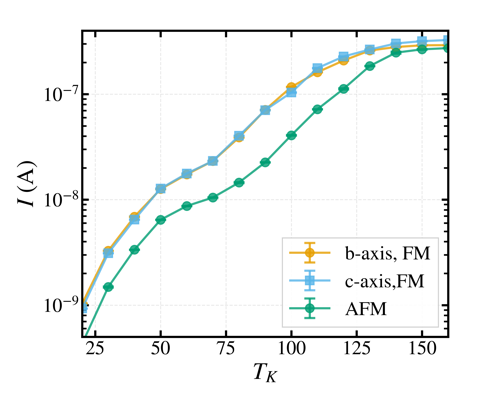}
    \caption{
    \textbf{Junction current at $V = \SI{0.5}{V}$ versus temperature.} Field-polarized FM state along the easy $b$-axis (orange) and the hard $c$-axis (blue), and the AFM ground state (green). The $b$- and $c$-axis FM currents coincide and both exceed the AFM current; counterpart of \autoref{fig:two_regimes}b.
    }
    \label{fig:S_baxis_IT}
\end{figure}

\autoref{fig:S_caxis_TMR} is the $c$-axis counterpart of \autoref{fig:TMR_bias_temperature}c,d: $\mathrm{MR}(V)$ is M-shaped at low temperature, with two peaks near $V \approx \pm \SI{0.5}{V}$, and the M-shape is lost toward $T_{\mathrm{N}}$ as on the $b$-axis. \autoref{fig:S_baxis_FN} is the counterpart of \autoref{fig:two_regimes}a: the $b$-axis AFM and FM states show the same Fowler--Nordheim character as the $c$-axis. \autoref{fig:S_baxis_IT} is the counterpart of \autoref{fig:two_regimes}b: the $b$- and $c$-axis field-polarized currents coincide across the whole temperature range and both exceed the AFM current, with the same activated-transport splitting on the two axes (Table~\ref{tab:S_arrhenius}).

\section{Angular law of the configuration-dependent band-edge offset}
\label{sec:S_model}

\textbf{Canting geometry.} For a magnetic field along the hard $c$-axis, the two sublattice magnetizations cant symmetrically toward the field by an angle $\alpha$ from the easy $b$-axis. Writing the reduced out-of-plane magnetization $m = M_{\mathrm{z}}/M_{\mathrm{sat}} = \sin\alpha$ and the reduced field $h = |H_{\mathrm{z}}|/H_{\mathrm{sat}}$, balancing interlayer exchange and Zeeman energy gives $\sin\alpha = h$ below saturation, so $m = h$. The two sublattice unit vectors are $\hat{\mathbf{s}}_{1} = \cos\alpha\,\hat{\mathbf{b}} + \sin\alpha\,\hat{\mathbf{c}}$ and $\hat{\mathbf{s}}_{2} = -\cos\alpha\,\hat{\mathbf{b}} + \sin\alpha\,\hat{\mathbf{c}}$, so the relative interlayer angle $\theta$ of the magnetizations obeys
\begin{align}
    \cos\theta = \hat{\mathbf{s}}_{1}\!\cdot\!\hat{\mathbf{s}}_{2} = 2m^{2} - 1 , \qquad \cos(\theta/2) = m = h ,
    \label{eq:S_geometry}
\end{align}
running from $\theta = \pi$ (AFM, $h=0$) to $\theta = 0$ (FM-aligned, $h=1$).

\textbf{Two candidate laws.} The two candidate angular laws for the offset are two perturbation orders of the same spin-dependent interlayer coupling $t_{\perp}(\theta) = t_{0}\cos(\theta/2)$ \cite{wilson2021interlayer, heienbuttel2024quadratic}. The quadratic law is the spin-filter alternating-barrier projection \cite{paudel2019spin, liu2022spin}, which weights the source-electrode spin onto the two sublattices and predicts
\begin{align}
    \Phi_{\mathrm{proj}}(\theta) = \Phi_{\mathrm{FM}} + \Delta\,\sin^{2}(\theta/2) = \Phi_{\mathrm{AFM}} - \Delta\,h^{2} ,
    \label{eq:S_projection}
\end{align}
quadratic in $h$ and flat at $h = 0$; the same $\sin^{2}(\theta/2)$ dependence is the second-order response reported for the bound exciton \cite{wilson2021interlayer, heienbuttel2024quadratic}. The linear law is the first-order response of the single-particle band edge: the coupling $t_{\perp}$ splits the degenerate adjacent-layer conduction-band-edge states in first order, lowering the bonding state by $|t_{\perp}|$, giving
\begin{align}
    \Phi_{\mathrm{hyb}}(\theta) = \Phi_{\mathrm{AFM}} - \left(\Phi_{\mathrm{AFM}} - \Phi_{\mathrm{FM}}\right)\cos(\theta/2) = \Phi_{\mathrm{AFM}} - |t_{0}|\,h ,
    \label{eq:S_hybridization}
\end{align}
linear in $h$, with $\Phi_{\mathrm{AFM}} - \Phi_{\mathrm{FM}} = |t_{0}|$ the bonding-state lowering at full FM alignment. The two laws differ at intermediate canting: \autoref{eq:S_projection} is flat at the AFM end and curves down, while \autoref{eq:S_hybridization} has a constant slope.

\begin{figure}[h]
    \centering
    \includegraphics[width = 0.6\linewidth]{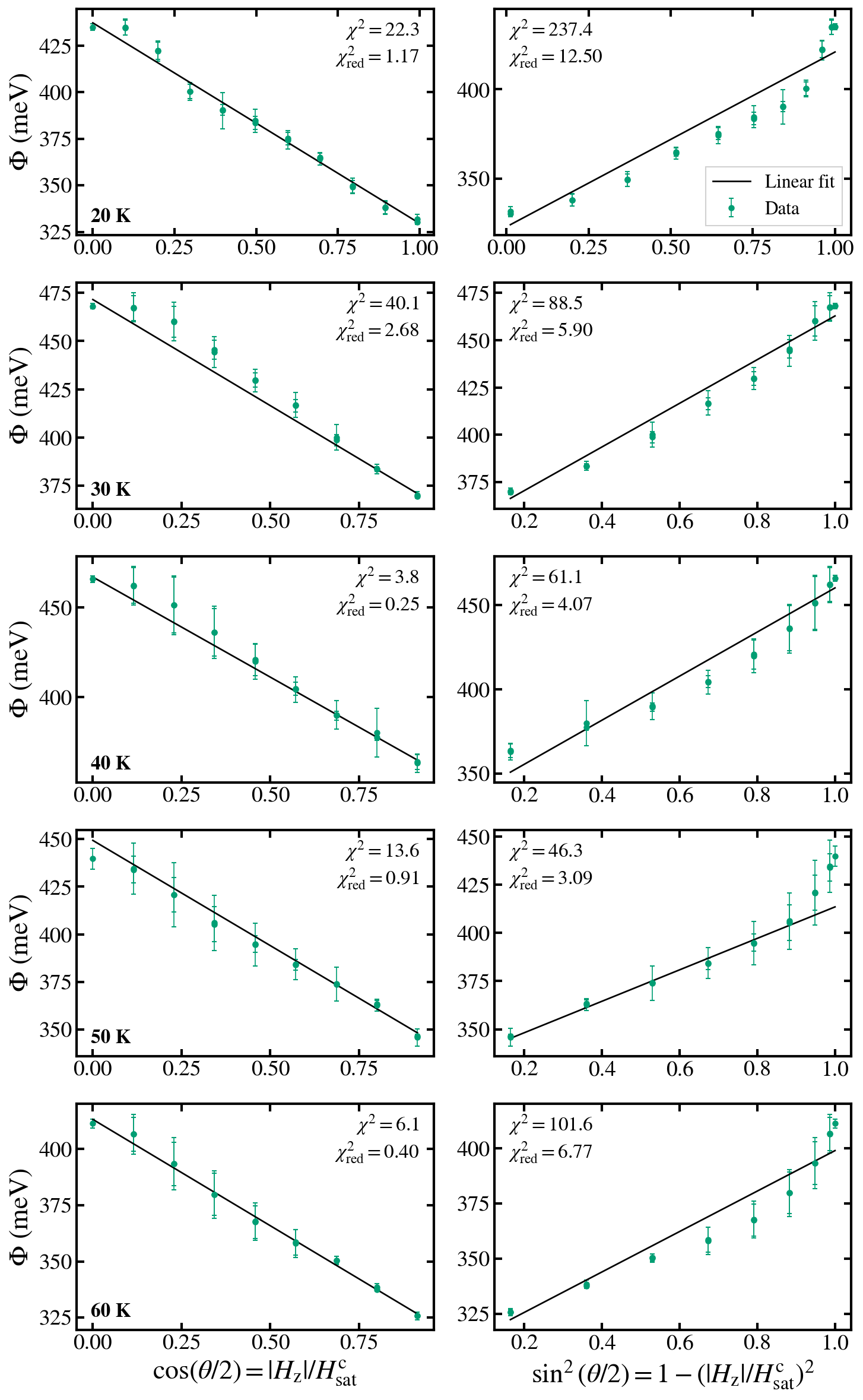}
    \caption{
    \textbf{Angular-law comparison of the configuration-dependent band-edge offset, \SI{20}{} to \SI{60}{K} ($c$-axis).} Each row is one temperature. \textbf{Left:} band-edge offset $\Phi$ versus $\cos(\theta/2) = |H_{\mathrm{z}}|/H_{\mathrm{sat}}$ with an inverse-variance-weighted linear fit, testing the first-order law (\autoref{eq:S_hybridization}). \textbf{Right:} $\Phi$ versus $\sin^{2}(\theta/2) = 1 - (|H_{\mathrm{z}}|/H_{\mathrm{sat}})^{2}$ with a linear fit, testing the second-order projection law (\autoref{eq:S_projection}). The $\chi^{2}$ and reduced $\chi^{2}$ are printed in each panel. The linear-in-$\cos(\theta/2)$ law fits with $\chi^{2}_{\mathrm{red}}$ of order unity, while the projection coordinate deviates systematically.
    }
    \label{fig:S_model_lin_sqr_lowT}
\end{figure}

\begin{figure}[h]
    \centering
    \includegraphics[width = 0.62\linewidth]{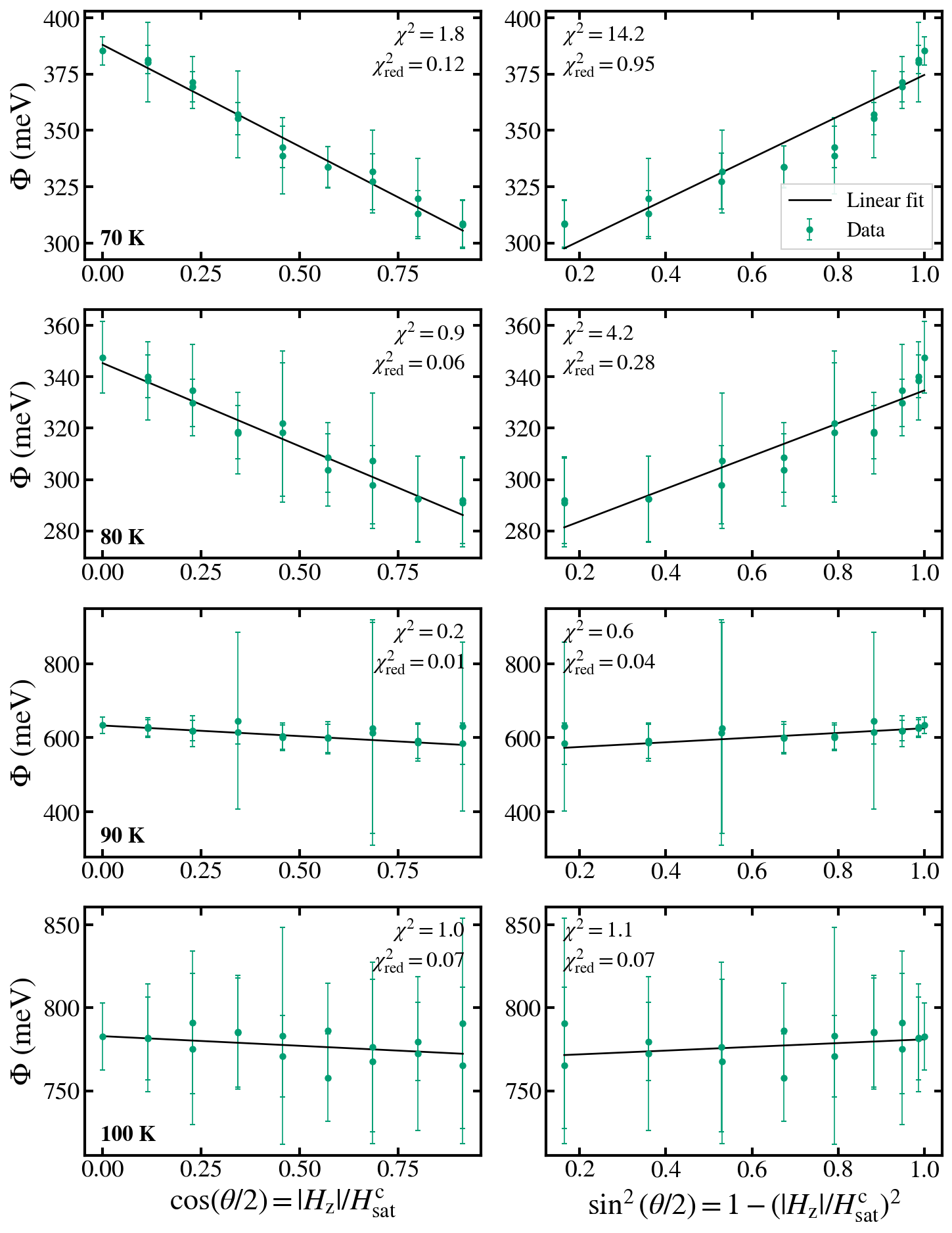}
    \caption{
    \textbf{Angular-law comparison, \SI{70}{} to \SI{100}{K} ($c$-axis).} Panels as in \autoref{fig:S_model_lin_sqr_lowT}. Above $\sim \SI{80}{K}$ the $\Phi(h)$ variation decreases as the field-emission regime is lost (Appendix~\ref{sec:S_fn}), both coordinates give $\chi^{2}_{\mathrm{red}} < 1$, and the two laws can no longer be distinguished.
    }
    \label{fig:S_model_lin_sqr_highT}
\end{figure}

\textbf{Model comparison.} We fit the binned below-saturation $\Phi(h)$ at each temperature with an inverse-variance-weighted straight line in each of the two coordinates, $\cos(\theta/2) = h$ (\autoref{eq:S_hybridization}) and $\sin^{2}(\theta/2) = 1 - h^{2}$ (\autoref{eq:S_projection}); the coordinate that linearises the data is identified by the smaller reduced $\chi^{2}$. From \SI{20}{} to \SI{60}{K} the $\cos(\theta/2)$ coordinate gives $\chi^{2}_{\mathrm{red}} = 0.25$ to $2.7$, against $3.1$ to $12.5$ for the projection coordinate (\autoref{fig:S_model_lin_sqr_lowT}, \autoref{tab:S_model}); the projection law has zero slope at $h = 0$ and does not reproduce the finite initial slope of the data. Above $\sim \SI{70}{K}$ the $\Phi(h)$ variation decreases as the field-emission regime is lost (Appendix~\ref{sec:S_fn}), both coordinates give $\chi^{2}_{\mathrm{red}} < 1$, and the two laws are no longer distinguishable (\autoref{fig:S_model_lin_sqr_highT}). The values $\Phi_{\mathrm{AFM}}$ and $\Phi_{\mathrm{FM}}$ themselves are model-independent, since at $\theta = \pi, 0$ the magnetic configuration is sharp.

\begin{table}[h]
    \centering
    \caption{Per-temperature reduced $\chi^{2}$ of the inverse-variance-weighted linear fit in each coordinate: $\cos(\theta/2)$ (first-order law, \autoref{eq:S_hybridization}) and $\sin^{2}(\theta/2)$ (second-order projection law, \autoref{eq:S_projection}), for the binned below-saturation $c$-axis $\Phi(h)$. The $\cos(\theta/2)$ coordinate is strongly preferred from \SI{20}{} to \SI{60}{K}; above $\sim \SI{70}{K}$ the variation decreases and the two become indistinguishable.}
    \label{tab:S_model}
    \begin{tabular}{l c c}
        \hline
        $T$ (K) & $\chi^{2}_{\mathrm{red}}$ [$\cos(\theta/2)$] & $\chi^{2}_{\mathrm{red}}$ [$\sin^{2}(\theta/2)$] \\
        \hline
        20  & 1.17 & 12.50 \\
        30  & 2.68 & 5.90  \\
        40  & 0.25 & 4.07  \\
        50  & 0.91 & 3.09  \\
        60  & 0.40 & 6.77  \\
        70  & 0.12 & 0.95  \\
        80  & 0.06 & 0.28  \\
        90  & 0.01 & 0.04  \\
        100 & 0.07 & 0.07  \\
        \hline
    \end{tabular}
\end{table}

\section{Bias-to-energy calibration and bias asymmetry}
\label{sec:S_pinning}

The normalized differential conductance $(\mathrm{d}I/\mathrm{d}V)/(I/V)$ peaks at $V_{\mathrm{peak}}$, which sits above the field-emission onset $V_{T}$ by the factor $V_{\mathrm{peak}} = c\,V_{T}$ with $c = 1.85$ measured at the AFM state (Appendix~\ref{sec:S_fn}), so the peak position gives the band-edge offset $\Phi = eV_{\mathrm{peak}}/c$. The graphite density of states is smooth across the bias window \cite{castroneto2009electronic}.

Two contributions separate the transport offset from the bare band-edge value. Additive, state-independent terms (interfacial charge transfer, image-charge lowering, and the work-function offset between the structurally inequivalent top and bottom \ce{FLG}/\ce{CrSBr} contacts) shift $\Phi_{\mathrm{AFM}}$ and $\Phi_{\mathrm{FM}}$ equally and cancel in the difference. The calibration factor is multiplicative and does not cancel: it rescales $\Phi_{\mathrm{AFM}} - \Phi_{\mathrm{FM}}$ by an order-unity factor. Comparisons with theory are therefore made at the level of scale. The same contact asymmetry gives an asymmetric Schottky barrier and the modest positive/negative asymmetry of $\mathrm{MR}(V)$ in the main text, as reported for \ce{CrI_3} \cite{song2018giant, song2019voltage} and \ce{EuS} \cite{miao2009magnetoresistance} junctions; the M-shape itself is symmetric in bias.

\section{Temperature dependence of the saturation field}
\label{sec:S_hsat}

The hard $c$-axis saturation field decreases with increasing temperature: we take $\mu_{0}H_{\mathrm{sat}} = \SI{2.2}{T}$ at base temperature and $\mu_{0}H_{\mathrm{sat}} \approx \SI{1.45}{T}$ in the \SI{90}{} to \SI{130}{K} window used for the high-temperature activation analysis. The reduced field $h = |H_{\mathrm{z}}|/H_{\mathrm{sat}}(T)$ uses the temperature-appropriate value throughout (Appendix~\ref{sec:S_model}, Appendix~\ref{sec:S_arrhenius}). The easier saturation near $T_{\mathrm{N}}$ is consistent with the dynamic magnetic crossover reported below $T_{\mathrm{N}}$ \cite{lopezpaz2022dynamic}.

\section{Differential-conductance temperature series}
\label{sec:S_dgdv}

\begin{figure}[h]
    \centering
    \includegraphics[width = \linewidth]{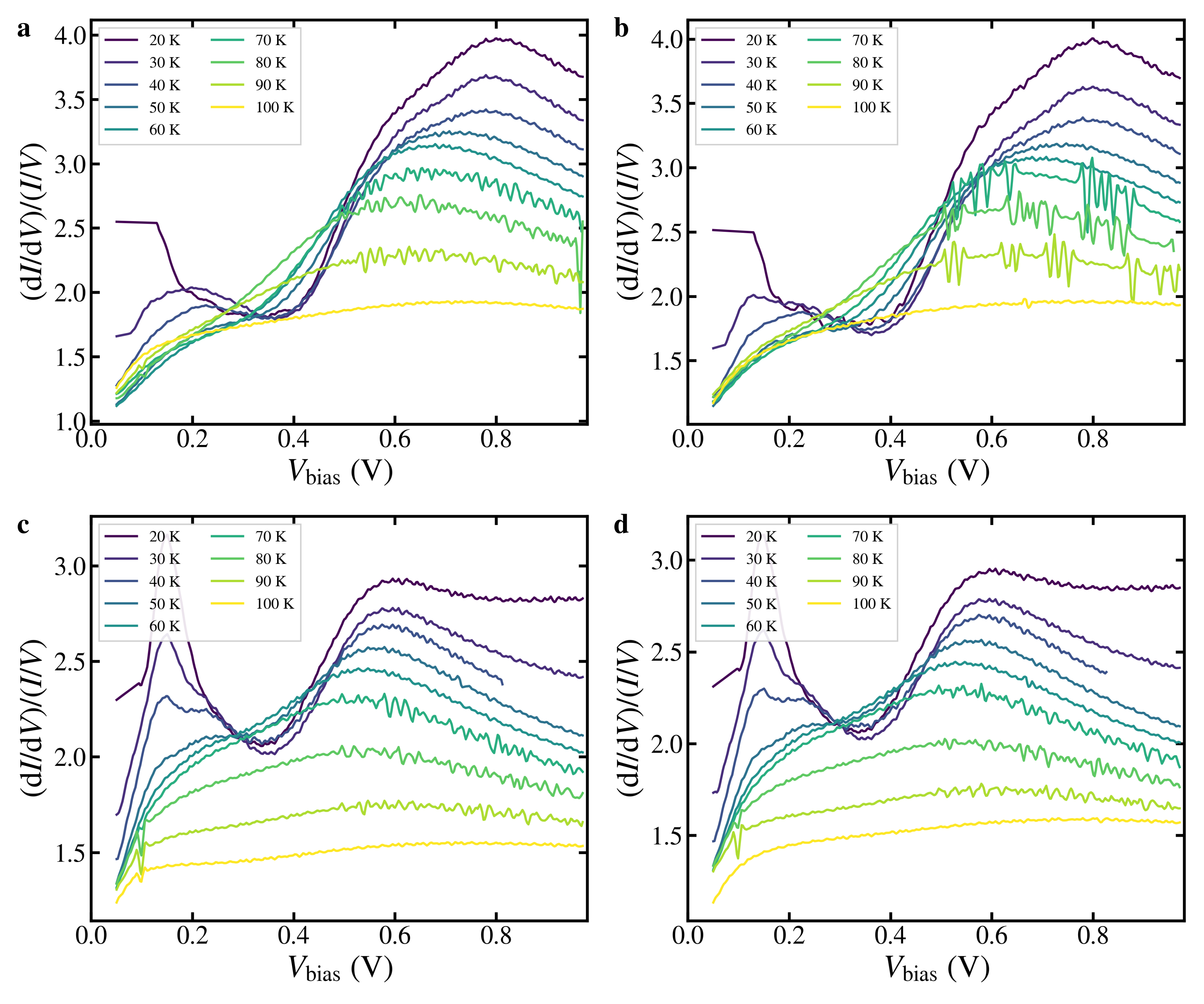}
    \caption{
    \textbf{Normalized differential conductance versus temperature.}
    \textbf{(a,b)} AFM state, $b$- and $c$-axis. \textbf{(c,d)} FM state, $b$- and $c$-axis. The band-edge-offset peaks of the FM and AFM states converge as $T \to T_{\mathrm{N}}$.
    }
    \label{fig:S_dGdV_temperature}
\end{figure}

\autoref{fig:S_dGdV_temperature} shows the normalized differential conductance $(\mathrm{d}I/\mathrm{d}V)/(I/V)$ in the FM and AFM states from \SI{20}{K} to \SI{100}{K} on both axes. The band-edge-offset peaks of the two magnetic states are well separated at low temperature and merge as $T \to T_{\mathrm{N}}$, following the decrease of the band-edge-offset difference that removes the magnetoresistance signal in the main text and mirroring the temperature decrease of the ARPES exchange splitting \cite{watson2024giant}.

\section{Fowler--Nordheim field-emission analysis: determining the band-edge offset}
\label{sec:S_fn}

\begin{figure}[h]
    \centering
    \includegraphics[width = \linewidth]{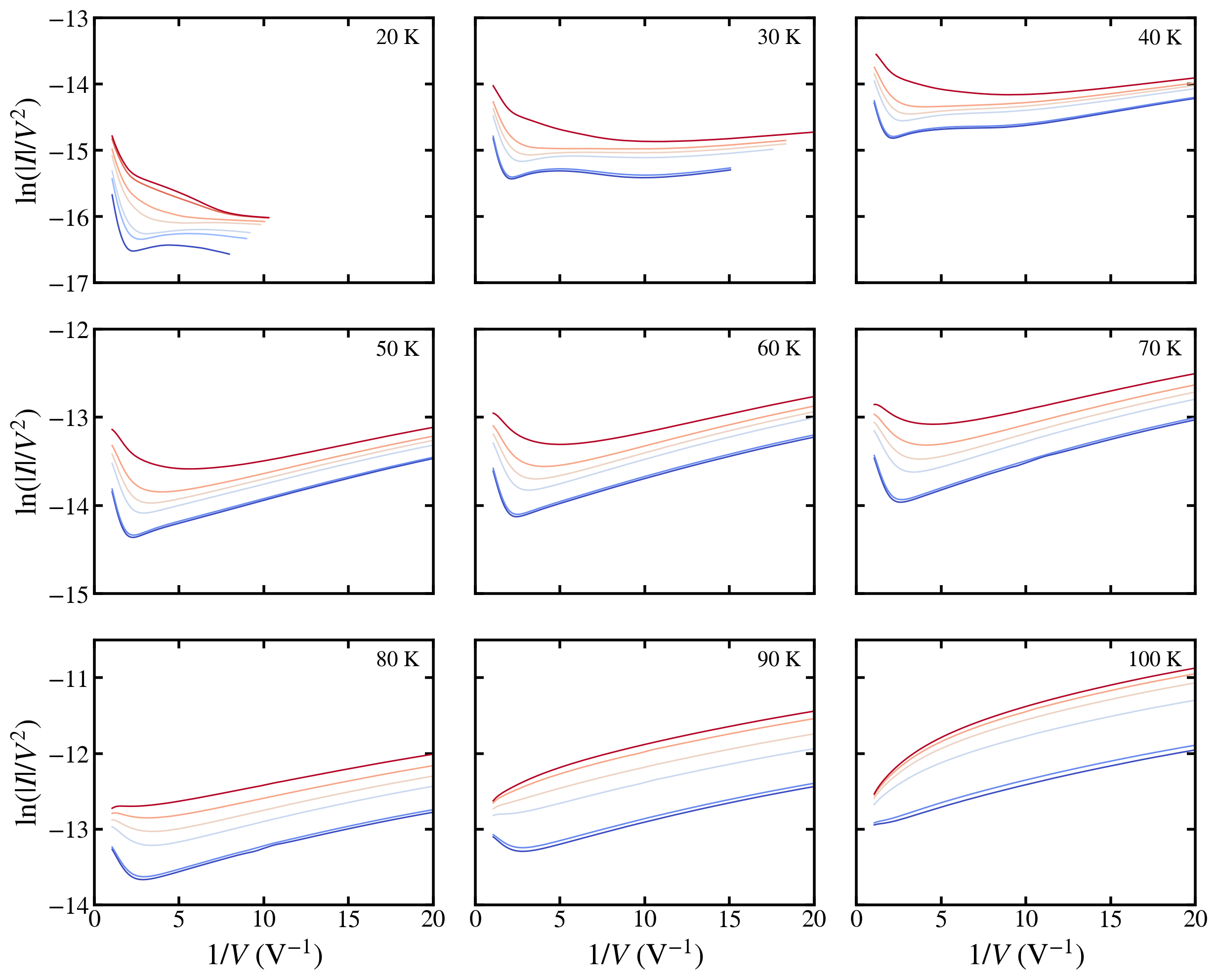}
    \caption{
    \textbf{Fowler--Nordheim representation versus temperature ($c$-axis).} $\ln(|I|/V^{2})$ versus $1/V$ for field-binned sweeps from the antiferromagnetic state ($H_{\mathrm{z}} = 0$, blue) to the field-polarized ferromagnetic state (saturated, red), one panel per temperature from \SI{20}{} to \SI{100}{K}. At low temperature the antiferromagnetic curve passes through a transition-voltage minimum, the crossover from low-bias hopping to field emission; the minimum fills in as the field cants the moments toward alignment and, with rising temperature, disappears entirely by $\sim \SI{90}{K}$ as the field-emission regime is lost.
    }
    \label{fig:S_FN_vs_T_c}
\end{figure}

\begin{figure}[h]
    \centering
    \includegraphics[width = \linewidth]{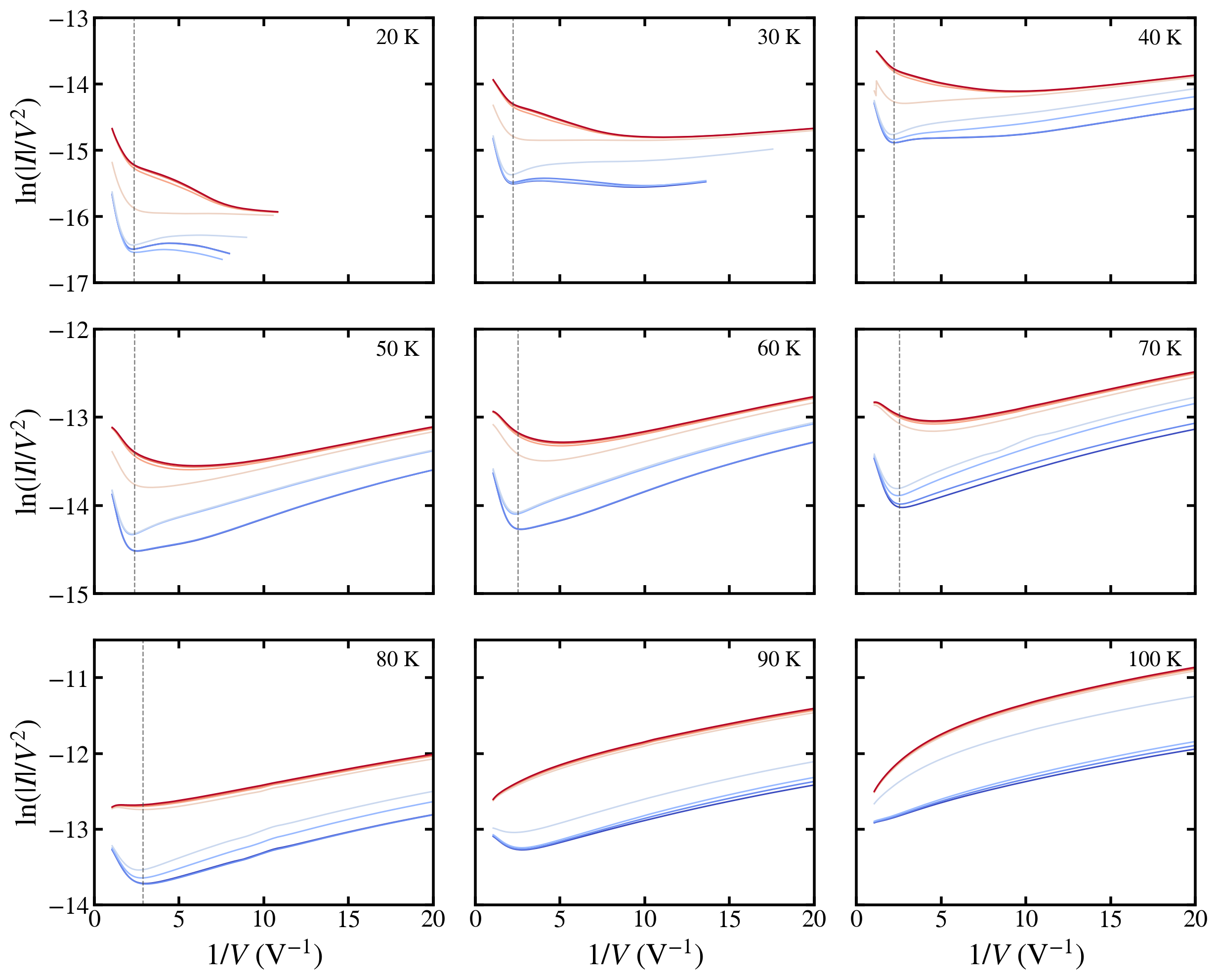}
    \caption{
    \textbf{Fowler--Nordheim representation versus temperature ($b$-axis).} As \autoref{fig:S_FN_vs_T_c}, for field-binned sweeps along the easy $b$-axis from the antiferromagnetic state (blue) to the field-polarized state (red); the dashed line marks the antiferromagnetic transition voltage $V_{T}^{\mathrm{AFM}}$. The $b$-axis AFM and FM states reproduce the field-emission character of the $c$-axis, indicating that the band-edge offset is a property of the magnetic state rather than of the field orientation used to reach it.
    }
    \label{fig:S_FN_vs_T_b}
\end{figure}

We analyze the high-bias transport in the Fowler--Nordheim (FN) representation $\ln(|I|/V^{2})$ versus $1/V$, field-binned from the antiferromagnetic state ($H = 0$) to the field-polarized ferromagnetic state, at each temperature from \SI{20}{} to \SI{100}{K} (\autoref{fig:S_FN_vs_T_c}, $c$-axis; \autoref{fig:S_FN_vs_T_b}, $b$-axis) \cite{fowler1928electron, simmons1963generalized}. As in the main text, the FN coordinate is used as a crossover diagnostic and as difference spectroscopy, not as textbook Fowler--Nordheim theory: at \SI{12}{nm} the current is hopping-dominated, and the narrow $c$-axis bandwidth invalidates the free-electron dispersion behind the textbook form \cite{lin2024influence}. In the antiferromagnetic state the curve falls linearly at high bias and passes through a minimum, the transition voltage $V_{T}$, which marks the crossover from low-bias hopping to field emission over the barrier \cite{beebe2006transition}. This minimum is the direct field-emission measure of the offset.

\textbf{Antiferromagnetic state calibration.} The two magnetic states behave differently in FN coordinates. In the antiferromagnetic state the adjacent layers are antiparallel, $\theta = \pi$, so the spin-dependent interlayer hopping amplitude $t_{\perp}(\theta) = t_{0}\cos(\theta/2)$ vanishes: the hybridization channel is shut, the hopping-to-field-emission crossover is abrupt, and the transition voltage is well resolved, $V_{T}^{\mathrm{AFM}} = \SI{0.435}{V}$ at \SI{20}{K} ($c$-axis, fit $R^{2} = 0.998$). In the field-polarized state $t_{\perp}$ is maximal, short-range interlayer hopping fills in the field-emission onset, and no transition-voltage minimum appears (\autoref{fig:S_FN_vs_T_c}). The calibration is therefore anchored in the antiferromagnetic state.

\textbf{Band-edge offset from the transition voltage and the calibration factor.} The transition voltage yields the offset, $eV_{T} \approx \Phi$ up to an order-unity calibration factor, independent of the barrier width $d$ and effective mass $m^{\ast}$ \cite{beebe2006transition}, so $\Phi_{\mathrm{AFM}} = eV_{T}^{\mathrm{AFM}} \approx \SI{0.44}{eV}$. The antiferromagnetic state is the only configuration providing both bias-scale observables, the conductance maximum $V_{\mathrm{peak}}^{\mathrm{AFM}}$ and the FN minimum $V_{T}^{\mathrm{AFM}}$, so the calibration factor is measured there:
\begin{align}
    c \equiv \frac{V_{\mathrm{peak}}^{\mathrm{AFM}}}{V_{T}^{\mathrm{AFM}}} = \frac{\SI{0.805}{V}}{\SI{0.435}{V}} = 1.85 .
    \label{eq:S_calibration}
\end{align}
This calibration is empirical, the measured ratio of two independent AFM-state observables; we do not derive it from transition-voltage-spectroscopy theory, whose quantitative predictions carry known model dependence \cite{trouwborst2011transition, baldea2012transition}. Every offset then follows from $\Phi = eV_{\mathrm{peak}}/c$. The ferromagnetic offset, whose FN minimum is not resolvable, is obtained by transferring the AFM-measured factor, $\Phi_{\mathrm{FM}} = eV_{\mathrm{peak}}^{\mathrm{FM}}/c = \SI{0.607}{V}/1.85 \approx \SI{0.33}{eV}$; applying $c$ to the FM state and to intermediate angles assumes it is configuration-independent. The band-edge-offset difference is
\begin{align}
    \Phi_{\mathrm{AFM}} - \Phi_{\mathrm{FM}} \approx \SI{0.11}{eV} ,
    \label{eq:S_barrier_split}
\end{align}
taken as the difference of the two calibrated offsets; the underlying conductance-peak separation is $\Delta V_{\mathrm{peak}} = \SI{0.193}{V}$. The additive, state-independent contributions cancel in this difference (Appendix~\ref{sec:S_pinning}), while the calibration factor rescales it without affecting its sign. The absolute values $\Phi_{\mathrm{AFM}} \approx \SI{0.44}{eV}$ and $\Phi_{\mathrm{FM}} \approx \SI{0.33}{eV}$ are indicative.

\textbf{Fowler--Nordheim slope.} The FN slope $B_{\mathrm{FN}} \propto m^{\ast 1/2}\Phi^{3/2} d$ mixes barrier height, width and effective mass, so the offset is not determined from it. At \SI{20}{K} the AFM and FM slopes are $B_{\mathrm{AFM}} = \SI{1.30}{V}$ and $B_{\mathrm{FM}} = \SI{0.66}{V}$; the rectangular-barrier slope ratio $(B_{\mathrm{AFM}}/B_{\mathrm{FM}})^{2/3} = 1.58$ shows only that the antiferromagnetic barrier is the larger of the two. Inverting the slope with $d = \SI{12}{nm}$ and a free-electron $m^{\ast}$ returns $\Phi \sim \SI{0.05}{eV}$, a factor of $\sim 6$ below the transition-voltage value, so the rectangular-barrier inversion does not describe this \SI{12}{nm} barrier; the transition-voltage determination is unaffected, being independent of $d$ and $m^{\ast}$. As a consistency check, $\ln(I_{\mathrm{FM}}/I_{\mathrm{AFM}})$ versus $1/V$ is linear with slope \SI{0.67}{V}, matching $B_{\mathrm{AFM}} - B_{\mathrm{FM}} = \SI{0.64}{V}$ to within $\sim 4\%$.

\textbf{Temperature range of validity.} The transition-voltage minimum is resolved only at low temperature. With rising temperature the ferromagnetic state leaves the field-emission regime first, its FN slope changing sign between \SI{70}{} and \SI{80}{K} and the antiferromagnetic slope between \SI{90}{} and \SI{100}{K}, by which point both curves rise monotonically (\autoref{fig:S_FN_vs_T_c}, \autoref{fig:S_FN_vs_T_b}). We therefore fix the band-edge offset at \SI{20}{} to \SI{30}{K} and carry the temperature trend with the differential-conductance peak splitting (Appendix~\ref{sec:S_dgdv}) and the activation analysis (Appendix~\ref{sec:S_arrhenius}).

\begin{table}[h]
    \centering
    \caption{Bias-scale observables, calibration factor and band-edge offsets at \SI{20}{K} ($c$-axis). The offset is determined directly from the antiferromagnetic transition voltage, $\Phi_{\mathrm{AFM}} = eV_{T}^{\mathrm{AFM}}$; the factor $c = V_{\mathrm{peak}}^{\mathrm{AFM}}/V_{T}^{\mathrm{AFM}}$, measured at the AFM state and assumed configuration-independent, transfers the calibration to the ferromagnetic state, $\Phi_{\mathrm{FM}} = eV_{\mathrm{peak}}^{\mathrm{FM}}/c$.}
    \label{tab:S_fn}
    \begin{tabular}{l c}
        \hline
        quantity & value \\
        \hline
        $V_{\mathrm{peak}}^{\mathrm{AFM}}$ (conductance maximum) & \SI{0.805}{V} \\
        $V_{\mathrm{peak}}^{\mathrm{FM}}$ (conductance maximum) & \SI{0.607}{V} \\
        $V_{T}^{\mathrm{AFM}}$ (FN transition-voltage minimum) & \SI{0.435}{V} \\
        $c = V_{\mathrm{peak}}^{\mathrm{AFM}}/V_{T}^{\mathrm{AFM}}$ (calibration factor) & $1.85$ \\
        $\Phi_{\mathrm{AFM}} = eV_{T}^{\mathrm{AFM}}$ (band-edge offset) & \SI{0.44}{eV} \\
        $\Phi_{\mathrm{FM}} = eV_{\mathrm{peak}}^{\mathrm{FM}}/c$ (band-edge offset) & \SI{0.33}{eV} \\
        $\Phi_{\mathrm{AFM}} - \Phi_{\mathrm{FM}}$ (band-edge-offset difference) & \SI{0.11}{eV} \\
        \hline
    \end{tabular}
\end{table}

\section{Two-regime activated transport and variable-range hopping}
\label{sec:S_arrhenius}

\textbf{Two regimes.} The current at a fixed low probe bias $V_{\mathrm{probe}} = \SI{0.2}{V}$, extracted per $I$--$V$ curve by a local linear fit and averaged over the AFM and FM field windows, shows two regimes in $\ln|I|$ versus $1/T$. A single Arrhenius fit over \SI{20}{} to \SI{160}{K} is curved; splitting it into a low-temperature regime (\SI{20}{} to \SI{80}{K}) and a high-temperature regime (\SI{90}{} to \SI{130}{K}, below $T_{\mathrm{N}}$), with a crossover gap excluded, resolves the two (\autoref{tab:S_arrhenius}). In the high-temperature regime the current is thermally activated over the band edge with $E_{\mathrm{a}}^{\mathrm{AFM}} \approx \SI{47}{meV}$, $E_{\mathrm{a}}^{\mathrm{FM}} \approx \SI{27}{meV}$, and an AFM-to-FM splitting $\Delta E_{\mathrm{a}} \approx \SI{20}{meV}$ on both axes. Lin \emph{et al.}\ report the same splitting for vertical hopping transport in \ce{CrSBr} in the ohmic regime ($E_{\mathrm{a}}^{\mathrm{AFM}} = 79 \pm 6$, $E_{\mathrm{a}}^{\mathrm{FM}} = 59 \pm 4$~meV) and attribute it to a conduction-band-edge downshift on AFM-to-FM switching \cite{lin2024influence}; their absolute activation energies differ from our finite-bias values.

\textbf{Bias robustness of the activated splitting.} Repeating the high-temperature fit for $V_{\mathrm{probe}}$ from \SI{0.05}{} to \SI{0.30}{V} lowers the absolute $E_{\mathrm{a}}$ of both states by $\sim \SI{15}{meV}$ through field-assisted barrier lowering of Poole--Frenkel or Schottky type, while the splitting $\Delta E_{\mathrm{a}}$ stays at \SI{18}{} to \SI{22}{meV} across the whole sweep, decoupling the magnetic splitting from the electrostatic barrier shape.

\textbf{Low-temperature variable-range hopping.} Below $\sim \SI{80}{K}$ the low-bias current is not activated over the band edge. Fitting the \SI{20}{} to \SI{80}{K} data to Arrhenius ($T^{-1}$), Mott ($T^{-1/4}$), and Efros--Shklovskii ($T^{-1/2}$) forms \cite{efros1975coulomb}, the Efros--Shklovskii (ES) law is preferred ($\Delta\mathrm{AIC} = 0$ versus $+7.31$ for Mott and $+9.40$ for Arrhenius), the signature of variable-range hopping through a Coulomb gap in disorder-localized band-tail states \cite{meir1996universal, huang2022conductivity}. Here $\sigma \propto \exp[-(T_{0}/T)^{1/2}]$ with $T_{0}$ set by the localization length and the density of states, not a gap, so the small apparent activation energy ($\sim \SI{6.5}{meV}$) obtained by forcing an Arrhenius fit is the local slope of a curved hopping line and is nearly identical for the two magnetic states ($\Delta E_{\mathrm{a}} \approx \SI{0.2}{}$ to $\SI{0.5}{meV}$). Nonetheless the chord-conductance magnetoresistance is large, \SI{130}{} to \SI{210}{\percent} at $V_{\mathrm{probe}} = \SI{0.2}{V}$.

\textbf{Hopping and the band-edge offset.} The VRH path does not involve the band edge, so its activation slope is insensitive to the band-edge shift of Appendix~\ref{sec:S_model}, while the same spin-dependent interlayer hopping $t_{\perp} = t_{0}\cos(\theta/2)$ sets the nearest-layer hopping amplitude, large in the FM state and suppressed in the AFM state. Lin \emph{et al.}\ report a shorter accessible hopping distance in the spin-aligned state, from two interlayer spacings to one \cite{lin2024influence}, appearing in their linear-regime data (\SI{10}{} to \SI{40}{K}) as a lowering of the ES parameter $T_{0}$; in our finite-bias data it appears instead as a field-dependent current magnitude at nearly field-independent slope.

\begin{figure}[h]
    \centering
    \includegraphics[width = \linewidth]{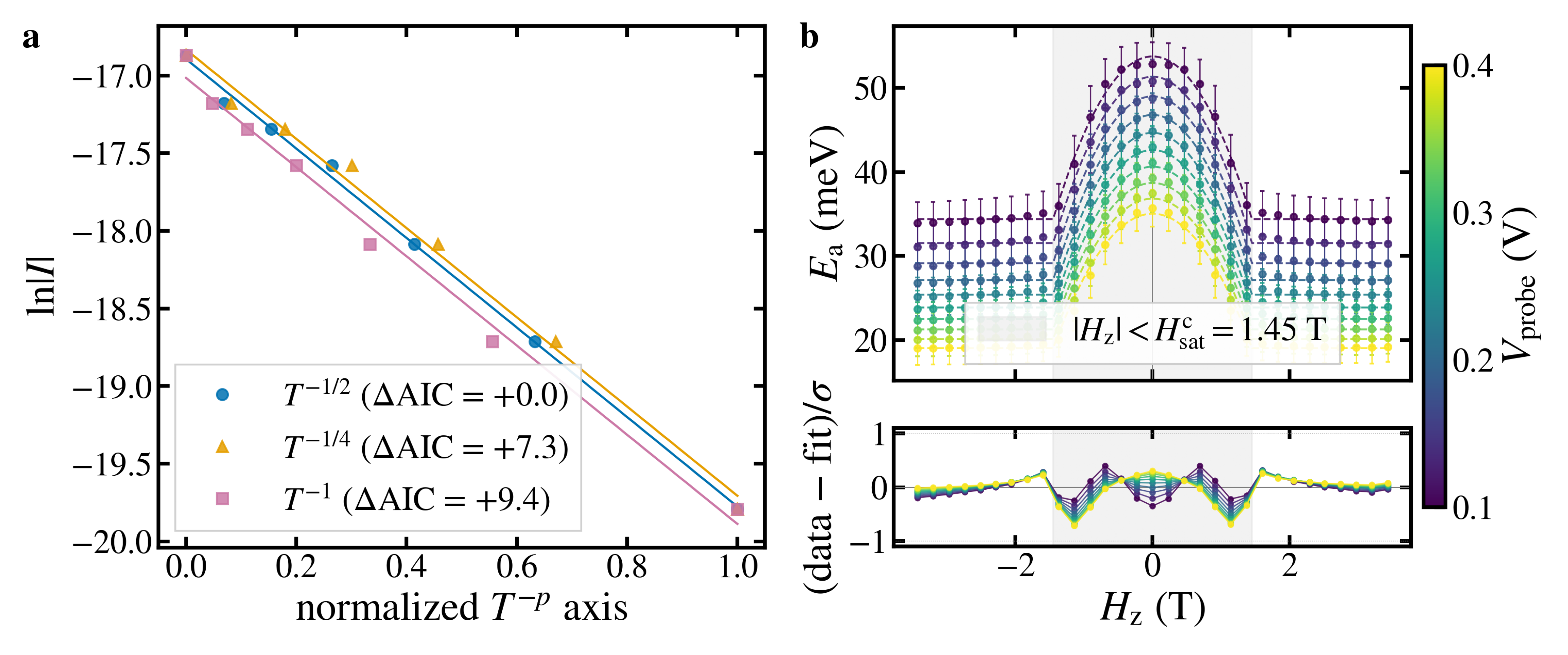}
    \caption{
    \textbf{Two-regime activated transport and variable-range hopping ($c$-axis).}
    \textbf{(a)} Low-temperature fit-form comparison: Efros--Shklovskii ($T^{-1/2}$) is preferred over Mott ($T^{-1/4}$) and Arrhenius ($T^{-1}$) by $\Delta\mathrm{AIC}$.
    \textbf{(b)} Activation energy $E_{\mathrm{a}}(H_{\mathrm{z}})$ in the high-temperature window for probe biases from \SI{0.10}{V} to \SI{0.40}{V} (color), with the spin-correlation model \autoref{eq:S_Ea_canting} (dashed); standardized residuals (lower panel) are within $\pm 1\sigma$ across all biases.
    }
    \label{fig:S_arrhenius_VRH}
\end{figure}

\textbf{Field dependence of the activation energy.} When a $c$-axis field cants the magnetizations in the high-temperature regime, the activation energy follows the interlayer spin correlation rather than the hybridization amplitude. The data obey
\begin{align}
    E_{\mathrm{a}}(m) = E_{\mathrm{a}}^{\mathrm{FM}} + \left(E_{\mathrm{a}}^{\mathrm{AFM}} - E_{\mathrm{a}}^{\mathrm{FM}}\right)\left(1 - m^{2}\right) ,
    \label{eq:S_Ea_canting}
\end{align}
with $m = M_{\mathrm{z}}/M_{\mathrm{sat}} = |H_{\mathrm{z}}|/H_{\mathrm{sat}}(T)$ and $H_{\mathrm{sat}} \approx \SI{1.45}{T}$ in this window. The interlayer spin correlation for canted sublattices is $\langle \mathbf{S}_{1}\!\cdot\!\mathbf{S}_{2}\rangle \propto \cos\theta = 2m^{2} - 1$ (\autoref{eq:S_geometry}), so its field-induced change from the AFM value is $\propto m^{2}$ and \autoref{eq:S_Ea_canting} is the form expected when the band edge follows the thermally averaged spin correlation. This form is adopted from the spin-disorder model of band-edge renormalization in magnetic semiconductors \cite{mauger1986magnetic}, not derived for \ce{CrSBr}; an s-d exchange coupling of the carriers to the magnetic order has been reported in \ce{CrSBr} for in-plane transport \cite{telford2022coupling}. It fits across the full probe-bias range (\SI{0.10}{} to \SI{0.40}{V}, \autoref{fig:S_arrhenius_VRH}b) with standardized residuals within $\pm 1\sigma$. \autoref{eq:S_Ea_canting} is mathematically the $\sin^{2}(\theta/2) = 1 - m^{2}$ form rejected for the low-temperature offset (Appendix~\ref{sec:S_model}): the field-emission channel responds to the hybridization amplitude (first order in $t_{\perp} \propto m$), the activated channel to the spin correlation ($\propto 2m^{2} - 1$).

\textbf{The crossover.} The transition between the two channels is a single crossover near \SI{90}{K}, $\sim 0.7\,T_{\mathrm{N}}$, appearing in three observables: the loss of the field-emission regime (Appendix~\ref{sec:S_fn}), the disappearance of the differential-conductance band-edge-offset peak (Appendix~\ref{sec:S_dgdv}), and the onset of the field-dependent activation energy here. It is the crossover that removes the M-shaped $\mathrm{MR}(V)$ in the main text.

\begin{table}[h]
    \centering
    \caption{Two-regime activation energies (meV) at $V_{\mathrm{probe}} = \SI{0.2}{V}$. The low-temperature regime is fit on \SI{20}{} to \SI{80}{K} and the high-temperature regime on \SI{90}{} to \SI{130}{K}; $\Delta E_{\mathrm{a}} = E_{\mathrm{a}}^{\mathrm{AFM}} - E_{\mathrm{a}}^{\mathrm{FM}}$.}
    \label{tab:S_arrhenius}
    \begin{tabular}{l l c c c}
        \hline
        axis & regime & $E_{\mathrm{a}}^{\mathrm{AFM}}$ & $E_{\mathrm{a}}^{\mathrm{FM}}$ & $\Delta E_{\mathrm{a}}$ \\
        \hline
        $c$ & low  & $6.60 \pm 0.28$  & $6.41 \pm 0.53$  & 0.2  \\
        $c$ & high & $46.79 \pm 2.56$ & $27.08 \pm 2.33$ & 19.7 \\
        $b$ & low  & $6.63 \pm 0.34$  & $6.10 \pm 0.51$  & 0.5  \\
        $b$ & high & $45.78 \pm 1.83$ & $24.29 \pm 1.28$ & 21.5 \\
        \hline
    \end{tabular}
\end{table}

\section{First-principles calculations}
\label{sec:S_dft}

\begin{figure}[h]
    \centering
    \includegraphics[width = \linewidth]{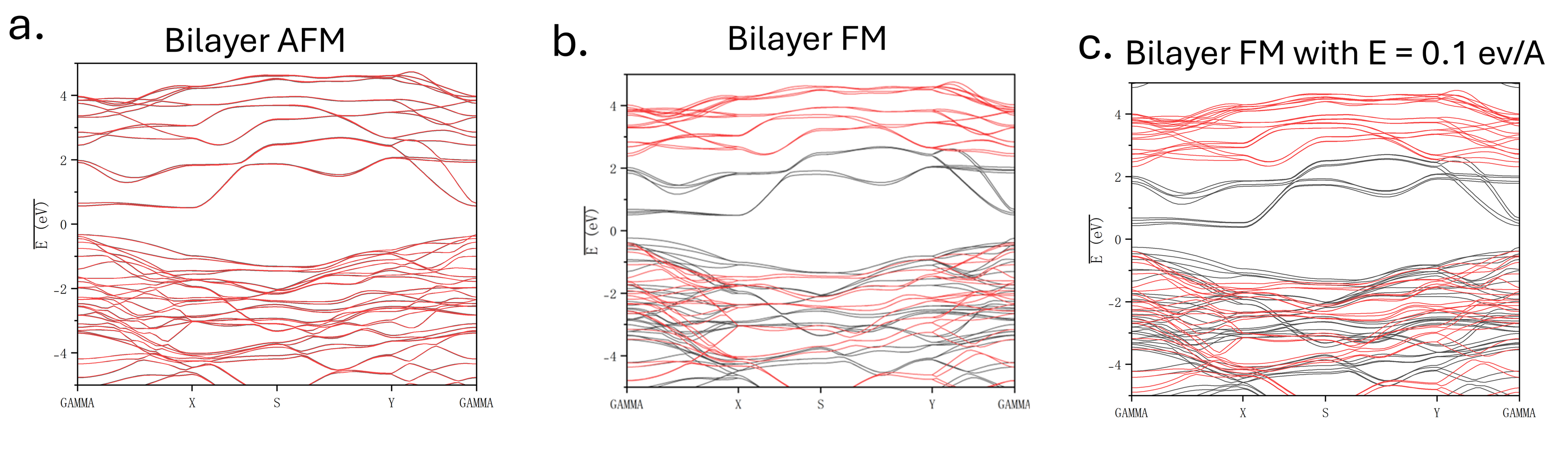}
    \caption{
    \textbf{Spin-resolved DFT on bilayer \ce{CrSBr}.} Spin-resolved band structures along $\Gamma$--X--S--Y--$\Gamma$ for \textbf{(a)} the AFM and \textbf{(b)} the FM configuration at zero applied field, and \textbf{(c)} the FM configuration under an applied out-of-plane electric field of \SI{0.1}{eV/\angstrom}. The two spin channels are drawn in black and red. At zero field the band gap is smaller in the FM configuration by $\sim\SI{0.12}{eV}$.
    }
    \label{fig:S_DFT_bilayer}
\end{figure}

Spin-resolved density-functional-theory calculations on bilayer \ce{CrSBr} give a zero-field band-gap reduction of $\sim\SI{0.12}{eV}$ on switching from the AFM to the FM configuration (\autoref{fig:S_DFT_bilayer}). That is the same $\sim\SI{0.1}{eV}$ scale as the band-edge-offset difference measured in transport. In this calculation the change sits almost entirely on the valence-band maximum, while the conduction-band-minimum eigenvalue shifts by only $\sim\SI{2}{meV}$. That partition between the two band edges is not reliable here: \ce{CrSBr} has two near-degenerate conduction-band minima at $\Gamma$ separated by $\sim\SI{40}{meV}$ \cite{wilson2021interlayer}, so the state labeled the conduction-band minimum can differ between the AFM and FM cells, which suppresses the apparent shift. Independent calculations do place a shift on the conduction band: $GW$ gives an electron interlayer coupling of \SI{29}{meV/T} at $\Gamma$ \cite{heienbuttel2024quadratic}, and density-functional calculations on multilayer \ce{CrSBr} give a conduction-band-edge downshift of $\sim\SI{50}{meV}$ on AFM-to-FM switching \cite{lin2024influence}. The main text therefore compares only the AFM-to-FM difference and its energy scale, which does not depend on how the calculated gap change divides between the two edges. The calculations also show that an out-of-plane electric field modulates the band gap and the band-edge positions (\autoref{fig:S_DFT_bilayer}c) \cite{guo2023electric, yao2025switching}, a further route by which the offset measured in transport can differ from the bare bilayer value.

\bibliography{../references}

\end{document}